%% file: JASA-template.tex
\documentclass[12pt]{article}
\usepackage{url} 

\usepackage{amsmath,amssymb,epsfig,epstopdf,titling,url,array,txfonts, csquotes, upgreek, physics, mathrsfs, mathtools, dsfont, caption, subcaption, rotating, float, graphicx,psfrag,epsf,enumerate, bbm, bm, arydshln, multirow, enumitem, authblk, verbatim}
\usepackage{natbib} 

\pdfoutput=1

\newcommand{\blind}{1}

\newcommand{\calT}{\mathcal{T}}
\newcommand{\calP}{\mathcal{P}}
\newcommand{\calH}{\mathcal{H}}
\newcommand{\calC}{\mathcal{C}}
\newcommand{\calR}{\mathcal{R}}
\newcommand{\calM}{\mathcal{M}}
\newcommand{\calG}{\mathcal{G}}

\newcommand{\calF}{\mathcal{F}}

\begin{document}

\def\spacingset#1{\renewcommand{\baselinestretch}%
{#1}\small\normalsize} \spacingset{1}

\if1\blind
{
  \title{\bf A Bayesian Dynamic Graphical Model for Recurrent Events in Public Health}
  
\author[1, *]{Aditi Shenvi}
\author[2, 3]{Jim Q. Smith}

\affil[1]{Centre for Complexity Science, University of Warwick, United Kingdom}
\affil[2]{Department of Statistics, University of Warwick, United Kingdom}
\affil[3]{The Alan Turing Institute, London, United Kingdom}
\affil[*]{Corresponding author: Aditi Shenvi; A.Shenvi@warwick.ac.uk}

\date{\vspace{-5ex}}

  \maketitle
} \fi

\if0\blind
{
  \bigskip
  \bigskip
  \bigskip
  \begin{center}
    {\LARGE\bf A Bayesian Dynamic Graphical Model for Recurrent Events in Public Health}
\end{center}
  \medskip
} \fi

\bigskip
\begin{abstract}
To analyze the impacts of certain types of public health interventions we need to estimate the treatment effects and outcomes as these apply to heterogeneous open populations. Dynamically modifying populations containing risk groups that can react very differently to changes in covariates are inferentially challenging. Here we propose a novel Bayesian graphical model called the Reduced Dynamic Chain Event Graph (RDCEG) customized to such populations. These models generalize the tree-based Chain Event Graphs to a particular class of graphically supported semi-Markov processes. They provide an interface between natural language explanations about what might be happening to individuals and a formal statistical analysis. Here we show how the RDCEG is able to express the different possible progressions of each vulnerable individual as well as hypotheses about probabilistic symmetries within these progressions across different individuals within that population. We demonstrate how well-developed Bayesian Network technologies can be transferred almost seamlessly to this class. Our work is motivated by the challenge of modeling non-pharmacological interventions for recurrent event processes. We illustrate our methodology in two settings: an intervention to reduce falls among the elderly and a trial to examine effects of deferred treatment among individuals presenting with early epilepsy.
\end{abstract}

\noindent%
\vfill

\newpage
\spacingset{1.45} 
\section{Introduction}
\label{sec:intro}
\input{Sections/Introduction.tex}

\section{The Reduced Dynamic Chain Event Graph}
\label{sec:rdceg}
\input{Sections/RDCEG.tex}

\section{Parameter Estimation and Model Selection}
\label{sec:parameter}
\input{Sections/Parameter.tex}

\section{Conditional Independence}
\label{sec:conditional}
\input{Sections/Conditional.tex}

\section{A Semi-Markov Representation}
\label{sec:semimarkov}
\input{Sections/Semimarkov.tex}

\section{A Simulation Study}
\label{sec:simulation}
\input{Sections/Simulation.tex}

\section{New Study of Epileptic Seizures}
\label{sec:case}
\input{Sections/Case.tex}

\section{Concluding Remarks}
\label{sec:discussion}
\input{Sections/Discussion.tex}

\subsection*{Acknowledgements}
The authors thank Tony Marson and David Chadwick for providing the MRC Multicentre Trial in Early Epilepsy and Single Seizures. The authors would also like to thank Jane Hutton and Laura Bonnett for their valuable inputs and suggestions for the epilepsy modelling.

\subsection*{Funding}
Jim Q. Smith was supported by the Alan Turing Institute and funded by the Engineering and Physical Sciences Research Council [grant number EP/K03 9628/1].

\bigskip
\begin{center}
{\large\bf SUPPLEMENTARY MATERIAL}
\end{center}

\section{Appendix A}
\label{appendixA}
\input{Appendices/AppendixA.tex}

\section{Appendix B}
\label{appendixB}
\input{Appendices/AppendixB.tex}

\section{Appendix C}
\label{appendixC}
\input{Appendices/AppendixC.tex}

\clearpage
\bibliographystyle{agsm}
\bibliography{biblio}

\end{document}

%% file: Sections/Introduction.tex
Here we describe and analyze a novel graphical model class called the reduced dynamic chain event graph (RDCEG). This is the first model of its kind to combine the technologies of the asymmetric chain event graphs (CEGs) and the flexible continuous time semi-Markov processes (SMPs), and is customized to modeling open populations. The RDCEG acquires several desirable properties from these two families. On the one hand, as a member of the CEG family, it can draw out functions of covariates important to explain responses of different subpopulations where the progress of individuals within the population can be highly heterogeneous \citep{collazo2018chain}. On the other, by importing technologies from SMPs, it is able to embed flexible holding times at each state and to handle irregularly sampled data \citep{barbu2009semi}. 

Because the RDCEG is graphical it shares many useful inferential properties with dynamic Bayesian networks. Like this class, the RDCEG unambiguously describes a family of statistical models. Furthermore, its graphical representation enables various conditional independence statements to be represented early in any analysis. In this way, expert structural domain knowledge can be embedded in the models and the implications of outputs of the statistical analysis fed back in an understandable way for scrutiny and criticism by the expert, see e.g. \cite{dawid2001separoids, pearl2009causality}. 

Because the likelihood of the model class factorizes, we can also show that the RDCEG class admits conjugate estimation and hence, fast model selection across different candidate models. In fact model selection and inference for this class can be performed completely analogously to Bayesian networks (BNs) and several well-developed BN technologies transfer almost seamlessly to this class. This transfer ensures that the broad implications of the RDCEG are transparent and that its statistical analysis is fast even within this highly multivariate heterogeneous domain. Note that, despite having imported BN technologies to the RDCEG, standard BN models could not be used directly to model these particular domains.

While the RDCEG can be applied in a wide range of applications including criminal behavior (see e.g. \citet{bunninbayesian2019}), for clarity we focus on its use to model recurrent events within public health; using it to assess the effects of non-pharmacological interventions. More specifically we are motivated by interventions designed to provide or improve services or access to treatments for the subpopulation that actively engages in their uptake. As in this context, data on certain vulnerable classes is sparse, we use a Bayesian approach for model selection and estimation so we can input prior information to regularize our technology. However, especially when there is a rich source of data this choice is not critical; other methods, for example penalized maximum likelihood selection could be used instead and very similar inferences ensue.
%

Depictions of open populations, i.e. populations where people can immigrate and emigrate, occur widely in ecology, conservation and epidemiology, see e.g. \cite{nisbet2003modelling, goffman1965epidemic}. The units of these population are in a constant state of flux due to a variety of reasons. This might include units moving from the region of study, their failing or improving health or death. In the contexts described in this article, although emigration causes missingness, the processes driving the missingness are not only difficult to correctly identify \citep{little2017conditions} but are often not random in any sense. Furthermore the population on which inference usually needs to be made is \textit{within} the current dynamically changing extant population and not the population from which this subpopulation is selected. For instance, incentivized pharmacies in Example 1 will typically be interested only in the responses of the subpopulation of smokers under their care. For these two reasons, within such a setting we find it expedient to build statistical models of the extant population directly rather than using the non-ignorable response methods which would necessarily involve an additional model of the, here very complex, missingness mechanism \citep{little2019statistical}. Hence, the RDCEG is parametrized conditioning on individuals \textit{not dropping out}. Effectively its topology and parametrization are chosen to correspond directly to measures of the existence of an effect and its extent on those who are active participants of the intervention. This is analogous to modeling the \textit{effect of the treatment on the treated} \citep{geneletti2011defining}.

Dynamic heterogeneous open populations come with an added challenge as they contain multiple subpopulations which evolve in very different ways in response to changes in their environments. In this setting we typically find that the covariates that measure such environments and can successfully explain the evolution of one subpopulation may have little or no impact on the evolution of another. Hence, it is often critical for our model to be able to hypothesize or discover these subpopulations in terms of different functions of their defining covariates. The structural forms of these functions that describe the different evolutions of such subpopulations is precisely what is represented within the RDCEG: see the illustration in Example 1.

It has long been recognized that SMPs are capable of identifying high-risk subgroups effectively (see e.g. \cite{hubbard2016using}). However, SMPs quickly become algebraically dense and so rather opaque. Despite this, this otherwise flexible class of models has sometimes been applied in the field of biomedicine and public health, see e.g. \cite{kang2006statistical, foucher2005semi, alaa2017learning}. Here we demonstrate how the RDCEG model can act as a vehicle to transfer this capability and flexibility of SMPs to a dynamic variant of the more tractable, expressive and interpretable graphical model family of CEGs. Each RDCEG has an SMP representation. Thus while retaining all the appealing properties of its parent graphical family, the RDCEG can be used to guide first passage and time to event analyses which do not typically form part of a graphical modeler's toolkit. Note that SMPs do have a graphical interface, namely its state transition diagram. However, this diagram is usually used not as a modeling tool in itself but rather as a pictorial description of some features of a chosen model. In \cite{barclay2015dynamic} it was shown that the colored versions of the state transition diagrams of a limited class of SMPs share many of the properties of dynamic bayesian networks. A major contribution of the RDCEG is that it formally generalizes this work so that it applies to discrete state, continuous time SMPs and then customizes it to models designed to study open populations. 

In Section \ref{sec:rdceg} we introduce the novel RDCEG class and its associated semantics. Here, for simplicity, we assume that all new entrants start from the same state and that the same RDCEG model is applicable to all these individuals irrespective of when they enter. Section \ref{sec:parameter} outlines conjugate Bayesian estimation, inference and model selection for this class. In Section \ref{sec:conditional}, we exploit the conditional independence structure depicted by these graphs to develop methods to query models, and in Section \ref{sec:semimarkov} we detail the construction of bespoke SMPs from an RDCEG depending on the results of such queries. In Sections \ref{sec:simulation} and \ref{sec:case} we proceed to demonstrate our methodology in two settings. The first example, based on a study by \cite{eldridge2005modelling} studies the impacts of a public health intervention concerning falls among the elderly. In the second example we apply our methodology to the MRC Multicentre trial for Early Epilepsy and Single Seizures dataset which was designed to examine the effects of immediate versus delayed treatment prescribed to individuals with early epilepsy. We conclude with a discussion on the future challenges of modeling with this class.

%% file: Sections/RDCEG.tex
\subsection{Notation}
\label{subsec:notation}

Event trees are a familiar step-by-step representation of how a process might evolve \citep{shafer1996art}. Here we extend the use of this term by annotating some of its edges with the time it takes to evolve from one state to another. In this way, an \textit{event tree} $\calT$ = ($V, E$) can be seen as a directed graphic representation of a continuous time process. Its possibly infinite vertex set is represented by $V$ and its edge set by $E$. The vertices represent the state occupied by an individual and the edges - ordered pairs of vertices - represent the transitional events that occur between these states. A directed path from vertex $v_0$ to vertex $v_k$ is a non-empty alternating sequence $v_0,e_{0,1},v_1,e_{1,2},\ldots,e_{k-1,k},v_k$ of vertices and edges such that $e_{i,i+1} = (v_i, v_{i+1}) \in E$ for all $i < k$. The unique vertex in $\calT$ with no edge entering it is called the \textit{root} $s_{0}$. The set of vertices with no emanating edges are called \textit{leaves}, denoted by $L(\calT)$ and the set of vertices which are not leaves are called \textit{situations}, denoted by $S(\calT)$. Variable $X_i$ on a situation $s_i$ denotes where the individual transitions next. It therefore has sample space $\mathbb{X}_i$ which is represented by the set of vertices into which edges emanating from $s_i$ enter. 

It is useful to partition the edge set $E$ into subsets $E^*$ and $E \backslash E^*$ where $E^*$ contains all edges which are assigned holding times. For instance, an edge depicting a transition from a healthy state to an ill state may have a holding time representing the duration for which the individual was healthy. Variable $H_{ik}$ denotes the holding time along the edge $e_{ik} \in E^*$ for a unit to have transitioned from situation $s_i$ to $s_k$.

In Appendix A of the supplementary material, we show how a probability measure can be defined on an infinitely large event tree. 

\subsection{The Definition of the RDCEG}

We now present a formal definition of the RDCEG class. The process of transforming an event tree into an RDCEG involves the construction of two intermediate colored graphs, namely the modified tree and the hued tree. While the RDCEG depicts events relevant to the extant population, we may choose to depict some \textit{critical terminating events} explicitly in the graph of the RDCEG. The choice of which dropout states (the set of which is represented by $L(\calT)$) can be considered as critical depends on the application and the purpose of modeling. Let $D^* \subseteq L(\calT)$ be the set of critical terminating events. In our depiction of the process, we can therefore remove the vertices $L(\calT) \backslash D^*$ and the edges entering them from the event tree to construct a \textit{modified tree} $\calM$. For instance, in Example 1 the terminating event of quitting smoking is of direct interest to modeling the uptake of smoking cessation services and so remains in the depiction, whilst dropouts related to individuals deregistering from the program, leaving the catchment area of the program provider or joining a different program are not depicted in the RDCEG. Henceforth, when we mention dropout events we refer to the events in $L(\calT) \backslash D^*$.

The transition probabilities along the edges emanating from each situation of $\calM$ are then renormalized to ensure that these probabilities sum to one for every situation. 

Next we partition the situations $S(\calM)$ into a set of \textit{stages} $\mathbb{U}$ and the edges in $E^*$ of the modified tree $\calM$ into a set of \textit{clusters} $\mathbb{C}$. These two partitions enable us to arrive at a finer partition of the situations called \textit{positions} $\mathbb{W}$. The partitions are identified as described below.

If there exists a bijection $\psi_u (s_i, s_j) : \mathbb{X}(i) \rightarrow \mathbb{X}(j)$ under which $X(i)$ and $X(j)$ can be hypothesized to share the same distribution and the edge labels correspond under the mapping, then situations $s_i$ and $s_j$ are said to be in the same stage $u \in \mathbb{U}$. Similarly, edges $e_{ij}$ and $e_{kl}$ where $i$, $j$, $k$ and $l$ need not be distinct, are in the same cluster $c \in \mathbb{C}$ if there exists a bijection $\psi_c' (e_{ij}, e_{kl}) : H_{ij} \rightarrow H_{kl}$ under which $H_{ij}$ and $H_{kl}$ have the same distribution. We then color the situations and edges in the modified tree to represent their stage and cluster memberships respectively. This transforms the modified tree into a \textit{hued tree} $\calH$. For simplicity, singleton stages and clusters, leaves, and edges belonging to $E \backslash E^*$ are uncolored. With its coloring, the hued tree graphically represents equivalences in one-step transition probabilities (through stages) and in holding times at situations before the next transition (through clusters).

Two situations $s_i, s_j \in u$, where $u \in \mathbb{U}$, are in the same position $w \in \mathbb{W}$ if and only if the hued trees $\calH_{i}$ and $\calH_{j}$ rooted at $s_i$ and $s_j$ are isomorphic. In this article, isomorphism refers to a structure, coloring and edge preserving mapping between two graphs. So two situations $s_i$ and $s_j$ in $S(\calH)$ are in the same position when the probability model of individuals arriving at either $s_i$ or $s_j$ are the same. The set of positions $\mathbb{W}$ along with a sink vertex $w_\infty$ - to collect the critically terminating events - form the vertex set of the RDCEG. Positions enable us to condense the infinitely large modified trees of the problems we consider into graphs with finite number of vertices and define the states of the implied SMP. 

\textit{Definition 1.} A \textit{Reduced Dynamic Chain Event Graph} $\calR = (V, E)$ is a directed colored graph with no self-loops. It is constructed from its associated hued tree $\calH$ by coalescing the situations belonging to the same position and collecting the leaves of the critical terminated trajectories into a single sink vertex $w_\infty$. The vertex set of $\calR$ is given by $V = \mathbb{W} \cup \{w_\infty\}$ where $\mathbb{W}$ is the set of positions of $\calH$ and between any two positions $w_i, w_j \in V$ there exist the same edges and edge labels as those between any $s_i \in w_i$ and $s_j \in w_j$ in $\calH$.

\begin{figure}[h!]
\centering
\begin{subfigure}[b]{0.40 \textwidth}
\centering
\includegraphics[trim = 0cm 0cm 0cm 0cm, scale = .175 ]{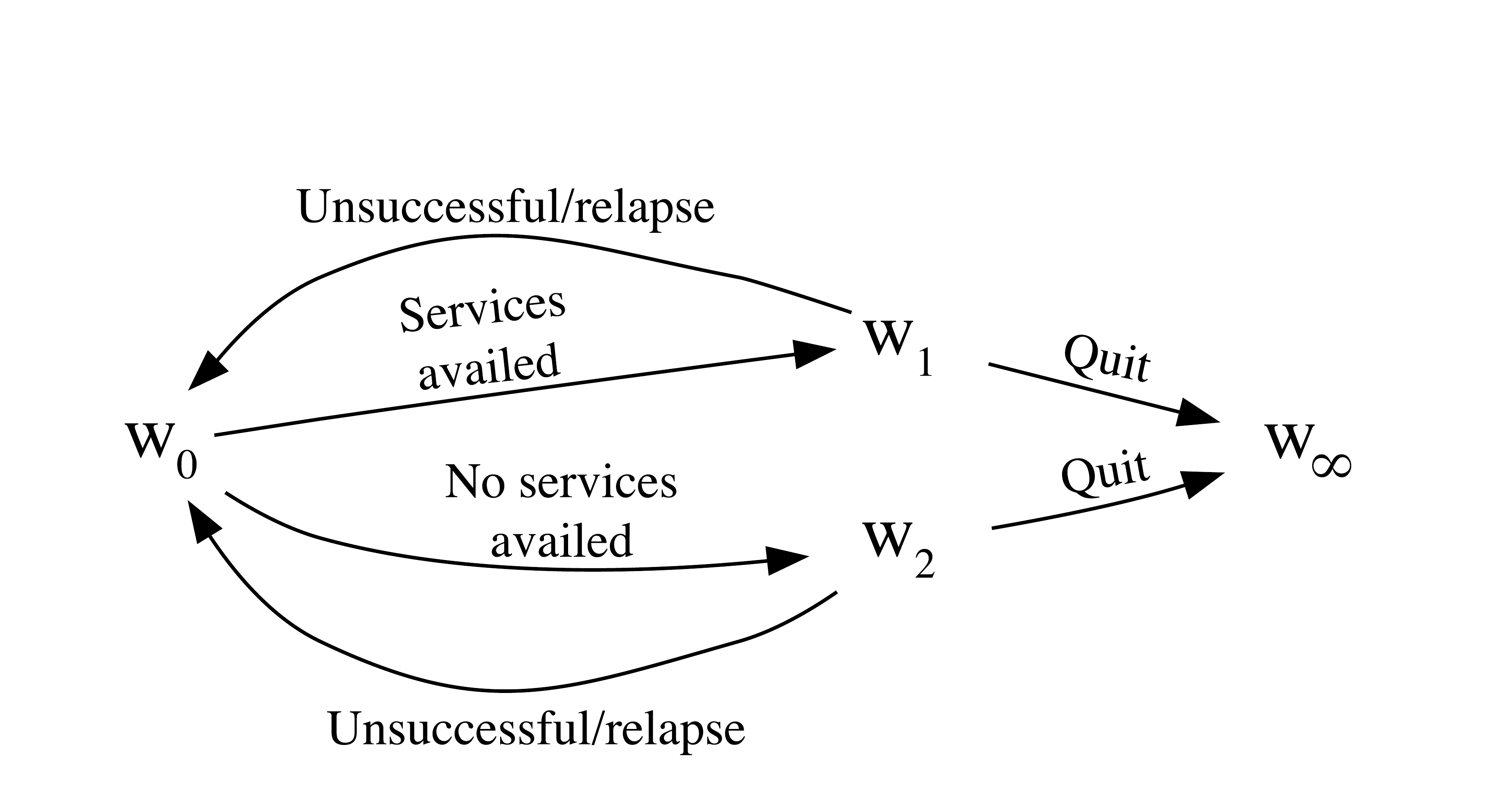}
\caption{}
\label{fig:smoking_rdceg_a}
\end{subfigure}
\begin{subfigure}[b]{0.40\textwidth}
\centering
\includegraphics[trim = 0cm 0cm 0cm 0cm, scale = .175 ]{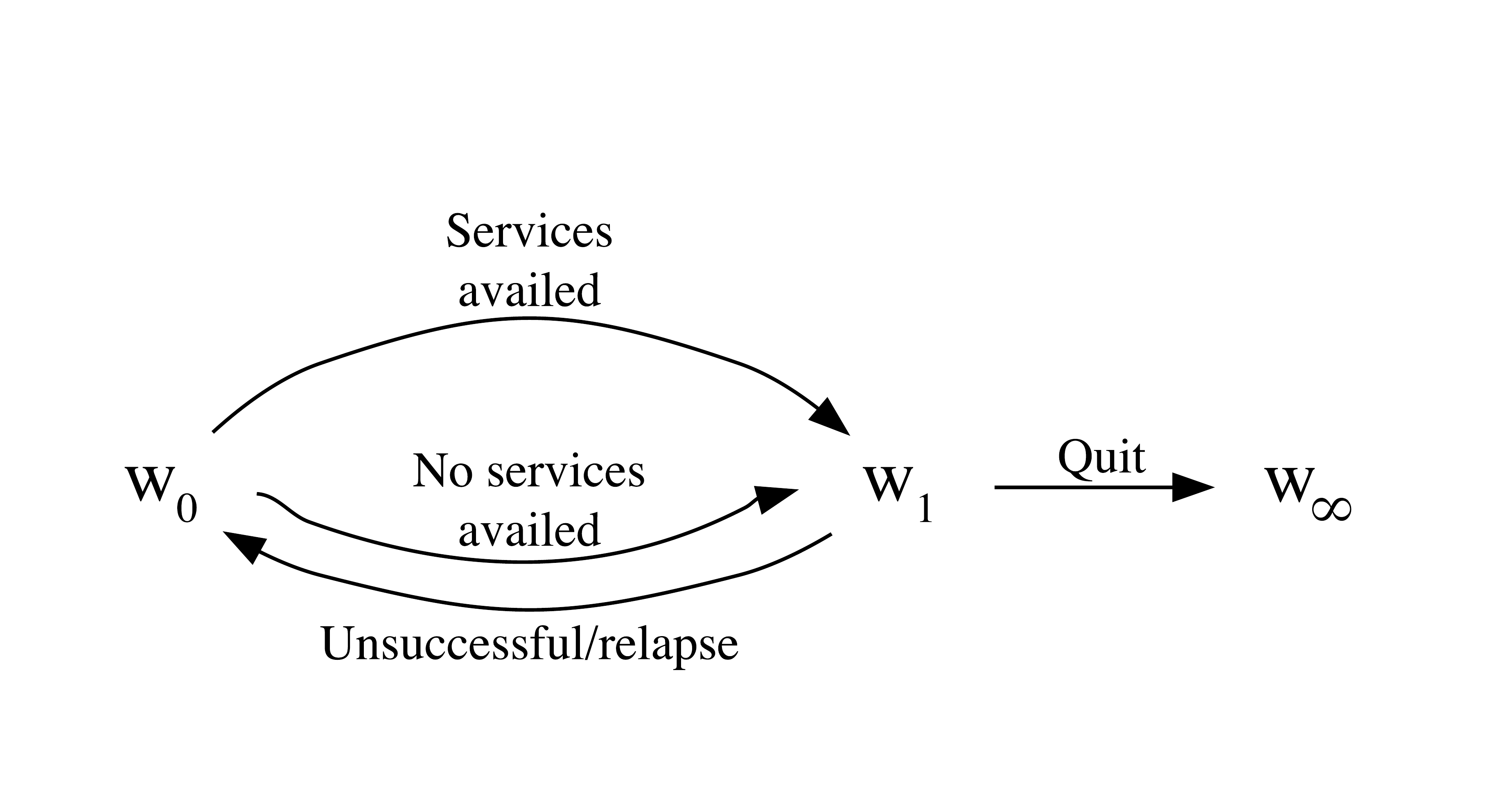}
\caption{}
\label{fig:smoking_rdceg_b}
\end{subfigure}
\caption{Competing RDCEG models for uptake of smoking cessation services.}
    \label{fig:smoking_rdceg}
\end{figure}

\textit{Example 1.} The two simple uncolored RDCEGs in Figure \ref{fig:smoking_rdceg} depict differing hypotheses for a process representing the uptake of cessation services provided to registered users of a smoking cessation program. Figure \ref{fig:smoking_rdceg_a} represents a difference in quitting outcomes based on utilization of the services while Figure \ref{fig:smoking_rdceg_b} hypothesizes that quitting is unaffected by whether or not the service was used. One advantage of the RDCEG over SMPs is already evident through this example. The state transition diagram of SMPs for both hypotheses is identical to the graph in Figure \ref{fig:smoking_rdceg_a} as SMPs do not allow for multiple edges in the same direction between two states. So this representation cannot distinguish these two very different hypotheses.

Edges in the RDCEG (such as $e_{1,0}$ and $e_{2,0}$ in Figure \ref{fig:smoking_rdceg_a} and $e_{1,0}$ in Figure \ref{fig:smoking_rdceg_b}) which represent a repetition of structure in its underlying hued tree are called \textit{cyclic edges}. A \textit{passage-slice} $\calP$ in an RDCEG $\calR$ describes the possible developments of an individual along the paths of $\calR$ after arriving at some specified situation $s_i$ up to the point at which it traverses a cyclic edge or arrives at the sink vertex. Passage-slices play the role of time-slices in the RDCEG. The first passage-slice $\calP_1$ in $\calR$ emanates from the root of $\calR$ and all the subsequent passage-slices $\calP_k$, $k = 2, 3, \ldots$, emanate from the vertices of $\calR$ into which the cyclic edges enter. For example, in Figures \ref{fig:smoking_rdceg_a} and \ref{fig:smoking_rdceg_b}, all the passage-slices are isomorphic to the entire graph of the RDCEG.

All examples used in this article have RDCEGs with time-homogeneous transitions and holding times which can be represented with a finite graph, and in which the holding time along an edge $e_{ij}$, representing a transition from situation $s_i$ to situation $s_j$, is independent of the probability of transitioning to situation $s_j$ from $s_i$.

%% file: Sections/Parameter.tex
In this section, we demonstrate how, with the appropriate mild assumptions, the likelihood of an RDCEG decomposes into two terms - one associated with its stages $\mathbb{U}$ and the other with its clusters $\mathbb{C}$. Here, we also set up estimation, inference and model selection for a homogeneous RDCEG.

\subsection{Parameter Estimation}
\label{subsec:parameter_estimation}

Consider an RDCEG $\calR = (V = \mathbb{W} \cup \{w_\infty\}, E)$ where $\mathbb{W}$ is the set of positions in the hued tree of $\calR$ and $w_\infty$ is the sink node if it exists. Each vertex $w \in V \backslash \{w_\infty\}$ represents a position. The conditional transition distribution $X_w$ on vertex $w$ is parametrized by $\boldsymbol{\pi}_w = (\pi_{w1}, \ldots, \pi_{wk_w})$, where $\{1,\ldots,k_w\}$ indexes the edges emanating from $w$. If edge $e_{wm} \in E^*$ (see Section \ref{subsec:notation}) where $e_{wm}$ is the $m$th edge emanating from $w$, then let its associated holding time distribution $H_{wm}$ be parametrized by $\theta_{wm}$. $\boldsymbol{\theta}_w = (\theta_{w1},\ldots,\theta_{wk_w})$ is the vector of holding time parameters associated with the emanating edges of $w$. For edges in $E \backslash E^*$, the holding time density can be set to 1 \textit{a priori} and \textit{a posteriori}. Here we assume that $\boldsymbol{\pi}_w$ and $\boldsymbol{\theta}_w$ are \textit{a priori} independent. This corresponds to the global parameter independence assumption in BNs \citep{spiegelhalter1990sequential} but adapted for this class. The joint density of moving along $e_{wm}$ after staying at vertex $w$ for time $h$ is
\begin{equation}
\mathbb{P}(e_{wm}, H_{wm} = h \,|\, w) \,=\,  \mathbb{P}(e_{wm}\,|\, w)\, \mathbb{P}(H_{wm} = h\,|\, e_{wm}, w) \,=\, \pi_{wm}\,f_{wm}(h),
\label{eq:joint_density}
\end{equation} 
\noindent where $f_{wm}$ is the holding time density function at $e_{wm}$. The joint density decomposes into separate terms for the densities of the conditional transition and holding time distributions. 

Suppose we have a random sample $X$ of $n$ individuals. Each individual $i$, $1 \leq i \leq n$ in the random sample traverses along the positions of the RDCEG $\calR$. We observe the sequence $\rho^i = (e_{w_{j_0},k_0}, h_{w_{j_0},k_0}, e_{w_{j_1},k_1}, \ldots, h_{w_{j_{n_i}},k_{n_i}})$ of edges and holding times associated with these edges for the path traversed by each individual $i$ where $w_{j_0} = w_0$ \citep{barclay2015dynamic}. Let $w_{j_l}$ be the $l$th position occupied by individual $i$, $e_{w_{j_l},k_l}$ the edge traversed after leaving $w_{j_l}$ and $h_{w_{j_l},k_l}$ the holding time at $w_{j_l}$ before this transition. Then, the likelihood of the probability vectors $\boldsymbol{\pi}$ and $\boldsymbol{\theta}$ is 
\begin{equation*}
L(\boldsymbol{\pi}, \boldsymbol{\theta} \,|\, \boldsymbol{N}, \boldsymbol{H}, \calR) \,=\, \prod_{i = 1}^{n} L(\boldsymbol{\pi}, \boldsymbol{\theta} \,|\, \rho^i, \calR) \, =\, \prod_{i = 1}^{n} \prod_{l = 1}^{n_i} \pi_{w_{j_l}{k_l}}^i \, f_{w_{j_l}{k_l}}^i(h_{w_{j_l}{k_l}}^i),
\label{eq:sample_likelihood} 
\end{equation*}
\noindent where \textbf{N} = $\{n_{wm}\}$, $n_{wm}$ denotes the number of times edge $e_{wm}$ has been traversed and \textbf{H} = $\{h_{wm}\}$, $h_{wm}$ is a vector of holding times on $e_{wm}$. Assuming that individual paths are independent given the parameters, we can therefore write
\begin{equation}
L(\boldsymbol{\pi}, \boldsymbol{\theta} \,|\, \boldsymbol{N}, \boldsymbol{H}, \calR) = \overbrace{\bigg \{ \prod_{w \in V \backslash \{w_\infty\}} \prod_{m = 1}^{k_w} \pi_{wm}^{n_{wm}} \bigg \} }^{L(\boldsymbol{\pi} \,|\, \boldsymbol{N}, \calR)} \overbrace{\bigg \{ \prod_{w \in V \backslash \{w_\infty\}} \prod_{m = 1}^{k_w} \prod_{l = 1}^{n_{wm}} f_{wm} (h_{wml}) \bigg\} }^{L(\boldsymbol{\theta} \,|\, \boldsymbol{N}, \boldsymbol{H}, \calR)},
\label{eq:stage_likelihood}
\end{equation}
\noindent where $h_{wml}$ is the holding time on edge $e_{wm}$ during the $l$th out of the $n_{wm}$ transitions. Thus the likelihood also decomposes over the parameters $\boldsymbol{\pi}$ and $\boldsymbol{\theta}$. It follows directly from the Bayes rule and the prior independence of the parameters that the impact of the data on $\boldsymbol{\pi}$ and $\boldsymbol{\theta}$ can be explored and modeled separately. 

Recall from Section \ref{sec:rdceg} that the set of positions $\mathbb{W}$ is a finer partition of the set of stages $\mathbb{U}$. Hence we can write $L(\boldsymbol{\pi} \,|\, \boldsymbol{N}, \calR) = \textstyle \prod\nolimits_{u \in \mathbb{U}}  \{ \textstyle \prod\nolimits_{m = 1}^{k_u} \pi_{um}^{n_{um}} \}$ where $k_u$ is the number of edges emanating from a situation in stage $u$. Similarly, $L(\boldsymbol{\theta} \,|\, \boldsymbol{N}, \boldsymbol{H}, \calR) = \textstyle \prod\nolimits_{c \in \mathbb{C}} \{ \textstyle \prod\nolimits_{l = 1}^{n_{c}} f_{c} (h_{cl}) \}$, where $\mathbb{C}$ is the set of clusters and $h_{cl}$ is the time spent in an edge in cluster $c$ during the $l$th out of total $n_{c}$ visits.

If interest lay in the full population then, the likelihood derived in Equation \ref{eq:stage_likelihood} is linked to the assumption of ignorable likelihood used when the missingness is at random and the parameters for the data generating process and the missingness mechanism are distinct (see e.g. \citet{little2019statistical}). However, in our setting, our model only concerns those who remain. This sidesteps the need for any assumptions about the missingness mechanism from an embedding superpopulation. Note that, therefore, the inferences from an RDCEG apply only to the population under study itself and to other populations with identical missingness mechanisms. 


\subsection{Bayesian Inference}
\label{subsec:inference}

Because of the likelihood decomposition (Equation \ref{eq:stage_likelihood}) it is possible to perform a conjugate analysis in this setting. So for example, Bayes Factor MAP based Bayesian model selection can be facilitated very quickly. Under the conjugate setting, we set the prior on the transition distribution parameter as $\boldsymbol{\pi}_u \overset{\text{ind}}{\sim} \text{Dir}(\alpha_u)$ where $\alpha_u$ is the Dirichlet concentration parameter for stage $u \in \mathbb{U}$. For simplicity we fix the shape parameter of the Weibull holding time distribution for each cluster $c \in \mathbb{C}$. The prior on its scale parameter $\theta_c$ is set as $\theta_c \overset{\text{ind}}{\sim} \text{IG}(\zeta_c, \beta_c)$ where $\text{IG}$ is the Inverse-Gamma distribution. The forms of the densities of the Dirichlet and Inverse-Gamma distributions are given in Section 1.7 of the supplementary material. Through a standard conjugate analysis, it follows that $\boldsymbol{\pi}_u | X \overset{\text{ind}}{\sim} \textmd{Dir}(\alpha^*_{u})$ where $\alpha_u^* = (\alpha_{u1}^*, \ldots, \alpha_{uk_u}^*)$ and $\theta_c | X \overset{\text{ind}}{\sim} \text{IG}(\zeta^*_{c}, \beta^*_c)$. The hyperparameters of $\alpha_{u}^*$ are a linear function of the prior and data given by $\alpha_{um}^* = \alpha_{um} + n_{um}$,  where $m = 1, \ldots, k_u$ and $\alpha_{um} = \alpha_u/ k_u$. The hyperparameters of $\theta^*_c$ are updated as $\zeta_{c}^* = \zeta_{c} + n_{c}$ and $\beta_{c}^* = \beta_{c} + \sum_{l = 1}^{n_{c}} (h_{cl})^{\kappa_{c}}$. 

A key advantage of the RDCEG compared to other dynamic graphical models such as the Dynamic Bayesian Network \citep{murphy2002dynamic} or the Continuous Time Bayesian Network \citep{nodelman2002continuous} is in its narrative. It describes the evolution of a process through events rather than random variables. Hence it does not need approximate inference methods to handle the problems arising due to temporal entanglement \citep{nodelman2002continuous}. Queries concerning the evolution of the process can be answered either by expressing the RDCEG as an infinite CEG (see Section \ref{sec:conditional}) or through the transition matrix of its SMP representation (see Section \ref{sec:semimarkov}).

\subsection{A Prior Specification for Model Selection}
\label{subsec:priors}

Typically priors are elicited from domain experts. In a modified tree, the prior specification of the Dirichlet hyperparameters can be set consistently across the different models by expressing these in terms of numbers of \textit{phantom units} the expert expects to see passing along the various edges of the tree \citep{freeman2011bayesian}. As the shape parameter of the Weibull distribution is known, information elicited about the moments of the scale parameter can be used to determine the hyperparameters of the Inverse-Gamma prior \citep{zhang2005bayesian}. Note that, we use the parametrized forms for these parameters due to the moderate size of our data. For Bayesian non-parametric analysis, see \cite{kottas2006nonparametric, hjort2010bayesian}. We could also use more sophisticated methods of prior specification (e.g. \citet{collazo2016new}). However at least for the examples used in this article we find that our model performs well even under a simple conjugate analysis.

Here, for illustrative purposes, we use weakly informative priors. We set the hyperparameters for $\boldsymbol{\pi}_{s}$ for situation $s \in S$ by treating them as phantom counts which is a straightforward adaptation from \cite{collazo2018chain}. For $\alpha$ phantom units starting at the root, if $\alpha'$ units reach situation $s$, then the hyperparameters at $s$ are $\alpha_{sm} = \alpha'/k$ for $\abs{\mathbb{X}(s)} = k$, $m = 1,\ldots,k$. The hyperparameter $\zeta_{sm}$ is set equal to $\alpha_{sm}$. We set the hyperparameter $\beta_{sm} = \tau_{sm}^{\kappa_{sm}}$ where $\kappa_{sm}$ is the known shape parameter of the associated Weibull distribution on edge $e_{sm}$ and $\tau_{sm}$ is the phantom waiting time at $s$ before going along edge $e_{sm}$. An alternative approach using the mean of the Inverse-Gamma distribution is presented in \cite{barclay2015dynamic}. 

\subsection{Model Selection for RDCEGs}
\label{subsec:model_selection}

Existing CEG Bayes Factor model selection methods \citep{freeman2011bayesian, cowell2014causal} are easily adapted to search across the much broader class of RDCEGs. As the parameter vectors $\boldsymbol{\pi}$ and $\boldsymbol{\theta}$ can be modeled independently under the conditions set above, we can perform hierarchical clustering separately on the situations and edges to obtain the set of stages $\mathbb{U}$ and set of clusters $\mathbb{C}$ respectively. From Equation \ref{eq:stage_likelihood}, the posterior density of the parameters separates as
\begin{equation*}
p(\boldsymbol{\pi}, \boldsymbol{\theta} \;| \boldsymbol{N}, \boldsymbol{H}, \calR) = p(\boldsymbol{\pi} \;| \boldsymbol{N}, \calR)\, p(\boldsymbol{\theta} \;| \boldsymbol{N}, \boldsymbol{H}, \calR)
\label{eq:posterior_density}
\end{equation*}
\noindent and the marginal likelihood of an RDCEG $\calR$ can be expressed in closed form as
\begin{equation}
 L(\calR \;|\boldsymbol{N}, \boldsymbol{H}) \,=\, L(\calR \;| \boldsymbol{N})\, L(\calR \;| \boldsymbol{N}, \boldsymbol{H})  \,=\, \prod_{u \in \mathbb{U}} \bigg \{ \dfrac{\Gamma(\sum_{m=1}^{k_u} \alpha_{um})}{\Gamma(\sum_{m=1}^{k_u} \alpha_{um}^*)} \; \prod_{m = 1}^{k_u} \dfrac{\Gamma(\alpha_{um}^*)}{\Gamma (\alpha_{um})} \bigg \}\;\times \, \prod_{c \in \mathbb{C}} \bigg \{ \dfrac{(\beta_{c})^{\zeta_{c}}}{\Gamma(\zeta_{c})} \; \dfrac{\Gamma(\zeta_{c}^*)}{(\beta_{c}^*) ^{\zeta_{c}^*}} \bigg \}.
\label{eq:marginal_likelihood}
\end{equation} 

Here, we focus on the log Bayes Factor $\log \textmd{BF} (\calR_1, \calR_2) =\log L(\calR_1 \;|\boldsymbol{N}, \boldsymbol{H}) - \log L(\calR_2 \;| \boldsymbol{N}, \boldsymbol{H})$ measuring the efficacy of a model $\calR_1$ compared to $\calR_2$ \citep{kass1995bayes}. Assuming all models are \textit{a priori} equally likely, this score is easily adapted to compare multiple models. Because $\log \textmd{BF}$ is linear in the components of the sets $\mathbb{U}$ and $\mathbb{C}$, $\log \textmd{BF}$ for two competing models would only include terms for the stages and clusters in which they differ. This greatly simplifies the process of model selection and can be exploited to develop fast and efficient model selection algorithms.

%% file: Sections/Conditional.tex
In this section we present the foundational theory of reading conditional independence statements directly from the graph of an RDCEG, extending analogous results for CEGs \citep{collazo2018chain, thwaites2015separation} to this new domain. The graph of an RDCEG, like that of a BN, encodes the dependence relationships between its variables that are implicit in its statistical model. Being constructed from a tree, it can also directly express \textit{context-specific} dependences that change depending on the realization of the variables. This is a key feature of the CEG family that other graphical families typically do not share \citep{boutilier1996context, geiger1996knowledge}. It has been repeatedly shown with respect to other graphical models (e.g. \cite{pearl2009causality}) that being able to read dependence relationships directly from the graph topology enhances the interpretability of each model within the class. It allows a form of structural diagnosis by enabling information to be read directly from the graph to the client for verification.


\subsection{Reading Conditional Independence from an RDCEG}
\label{subsec:ci_rdceg}
 
We state the following definitions for any directed acyclic graph $\calG = (V, E)$ which is colored according to its stage and cluster memberships. 

\textit{Definition 1.} Consider a set of vertices $U \subseteq V$ in $\calG$ such that if there exists a vertex of color $c$ in $U$ then all vertices of color $c$ in $V$ are in $U$. The set $U$ is called a \textit{cut} if all root-to-sink paths in $\calG$ pass through exactly one $u \in U$. 

\textit{Definition 2.} A set of vertices $W \subseteq V$ in $\calG$ is called a \textit{fine cut} if all root-to-sink paths in $\calG$ pass through exactly one $w \in W$.

Hence, a cut gives us a subset of stages and a fine cut gives us a subset of positions. We now state Theorem 1 which allows us to read conditional independence statements directly from the graph of an RDCEG. 

\textit{Theorem 1.} Consider a set of vertices $V'$ in the passage-slice $\calP_k$, $k \in \mathbb{N}$ of an RDCEG $\calR$. If $V'$ depicts a cut in $\calP_k$, given the stage $u \in V'$ occupied by an individual, the stage into which (s)he transitions from $u$ in $\calP_k$ is independent of his/her path into $u$. Similarly, if $V'$ represents a fine cut in $\calP_k$, given the position $w \in V'$ occupied by the individual, his/her future evolution in $\calP_{k+n}$, $n \in \mathbb{N}$ after leaving $w$ is independent of his/her path into $w$.

Theorem 1 enables us to query dependence relationships in an RDCEG concerning events stretching across $n$ passage-slices. The proof of this theorem involves the transformation of an RDCEG into its associated \textit{infinite CEG} rolled up to $n$ passage-slices denoted by $\calC_n$. The proof of Theorem 1, along with details of the construction of an infinite CEG, are presented in Appendix A of the supplementary material. So just as for the BN, the graph alone can be used to perform model validation.

As an RDCEG $\calR$ does not depict dropout events belonging to $L(\calT) \backslash D^*$ in its underlying event tree $\calT$, the conditional independence statements read from a cut or fine cut are implicitly conditional on the individual not dropping out due to an event in $d \in L(\calT) \backslash D^*$ at the next transition or in the next $n$ passage-slices respectively.

We now look at a brief example of the use of this theorem in practice.

\begin{figure}[h!]
\centering
\includegraphics[trim = 0cm 2cm 0cm 2cm, scale = .3 ]{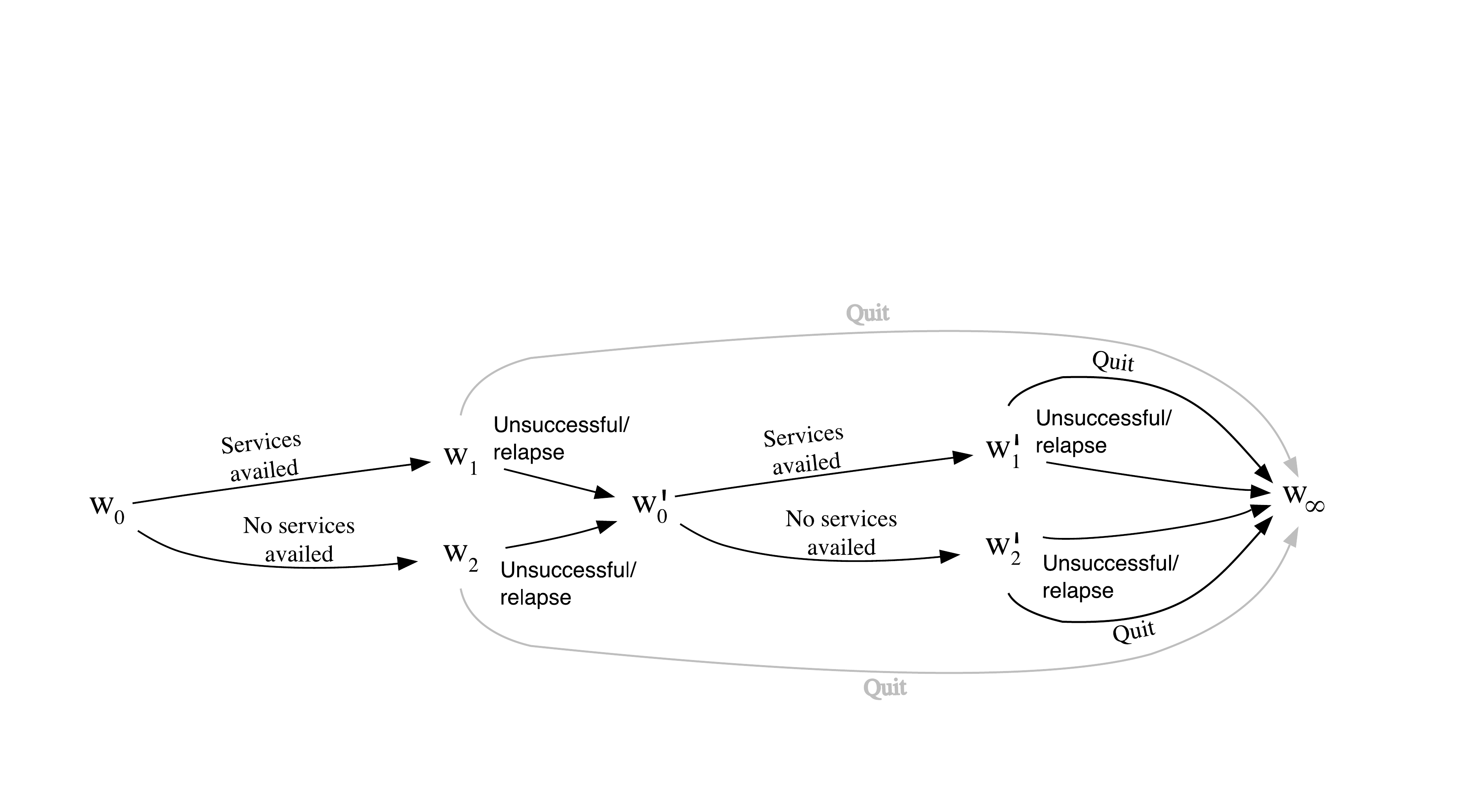}
\caption{$\calC_2$ of the RDCEG in Figure \ref{fig:smoking_rdceg_a}. The darkened edges represent the subgraph of $\calC_2$ from which we read the conditional independence statement described in the example.}
    \label{fig:rolled_out_ceg}
\end{figure}

\textit{Example 1 (continued).} In a study of registered users of a smoking cessation program, experts have hypothesized the RDCEG in Figure \ref{fig:smoking_rdceg_a} (its associated infinite CEG rolled out to two passage-slices $\calC_2$ is shown in Figure \ref{fig:rolled_out_ceg}). All paths of its first passage-slice pass through the subset $V' = \{w_1, w_2\}$ exactly once. The vertex occupied by a man along $V'$ indicates whether he used the cessation services in his last attempt to stop smoking. Once we know whether he is at $w_1$ or $w_2$, we can determine his probability of quitting at this attempt. Further, one consequence of this hypothesis is that if he has not quit at this attempt, so long as he does not dropout before his next attempt, his choice of using the services in his next attempt and the success of that attempt are not dependent on whether he availed these services in his previous attempt(s). From Theorem 1, we can therefore conclude that the model hypothesized by this graph implicitly asserts that there is no negative influence on uptake of services for those who availed the cessation services in past attempts but failed to quit smoking. This information can be fed back to the domain expert for verification and the model adapted if this deduction appears implausible.

%% file: Sections/Semimarkov.tex
Typically, graphical models are not equipped to answer queries related to first passage times and recurrences in the system, particularly when we introduce flexible holding times on the states. It is therefore useful at this stage to relate our methods to the class of semi-Markov representations of a given problem. The graph of the RDCEG - first queried as illustrated in the last section - relates to the state transition diagram for its representative SMP, enabling us to harness methodologies associated with SMPs \citep{barbu2009semi}. For a Bayesian treatment of SMPs, see e.g. \cite{butler2000bayesian}.

Recall that an SMP has two simultaneously evolving sub-processes. One concerns the state occupied by an individual, the other the time spent in each state. Consider a stochastic process \textbf{Z} = $\{Z_{t}, t \geq 0\}$ on a discrete state space \textbf{S}. The state occupied at the $n$th transition is given by \textbf{X} = $(X_{n})_{n \in \mathbb{N}}$, the jump times by \textbf{T} = $(T_{n})_{n \in \mathbb{N}}$ and the holding time in $X_n$ before moving to $X_{n+1}$ by $\boldsymbol{\tau}$ = $(\tau_{n})_{n \in \mathbb{N}}$ where $\tau_n = T_n - T_{n-1}$. The process \textbf{Z} is an SMP when
\begin{equation*}
\mathbb{P}(X_{n+1} = j, \tau_{n+1} \leq t  | X_{n},\ldots,X_{0}; \tau_{n},\ldots,\tau_{1}) = \mathbb{P}(X_{n+1} = j, \tau_{n+1} \leq t | X_{n} = i), \;i, j \in \textmd{\textbf{S}}, n \geq 1, t \geq 0.
\label{eq:semi-Markov}
\end{equation*}

The process \textbf{X} is called the \textit{embedded Markov chain} with transition probability matrix $P = (p_{ij})$ where $p_{ij} = \mathbb{P}(X_{n+1} = j\,|\, X_n = i)$. An SMP is completely defined by its renewal kernel $Q(t) = [Q_{ij}(t)| i,j \in \textbf{S}]$ and its initial distribution $p = [p_i| i \in \textbf{S}]$ where $p_i = \mathbb{P}(X_0 = i)$. The $ij$th entry of the renewal kernel $Q$ is 
\begin{equation}
Q_{ij}(t) = \mathbb{P}(X_{n+1}= j, \tau_{n+1} \leq t \,|\, X_n = i) = p_{ij}F_{ij}(t).
\label{eq:rewritten_renewal_kernel}
\end{equation}
\noindent where $F_{ij}(t) = \mathbb{P}(\tau_{n+1} \leq t \,|\, X_{n+1} = j, X_n = i)$ is the cumulative conditional holding time.

\begin{figure}[h!]
\centering
\includegraphics[trim = 0cm 2cm 0cm 2.85cm, scale = .15 ]{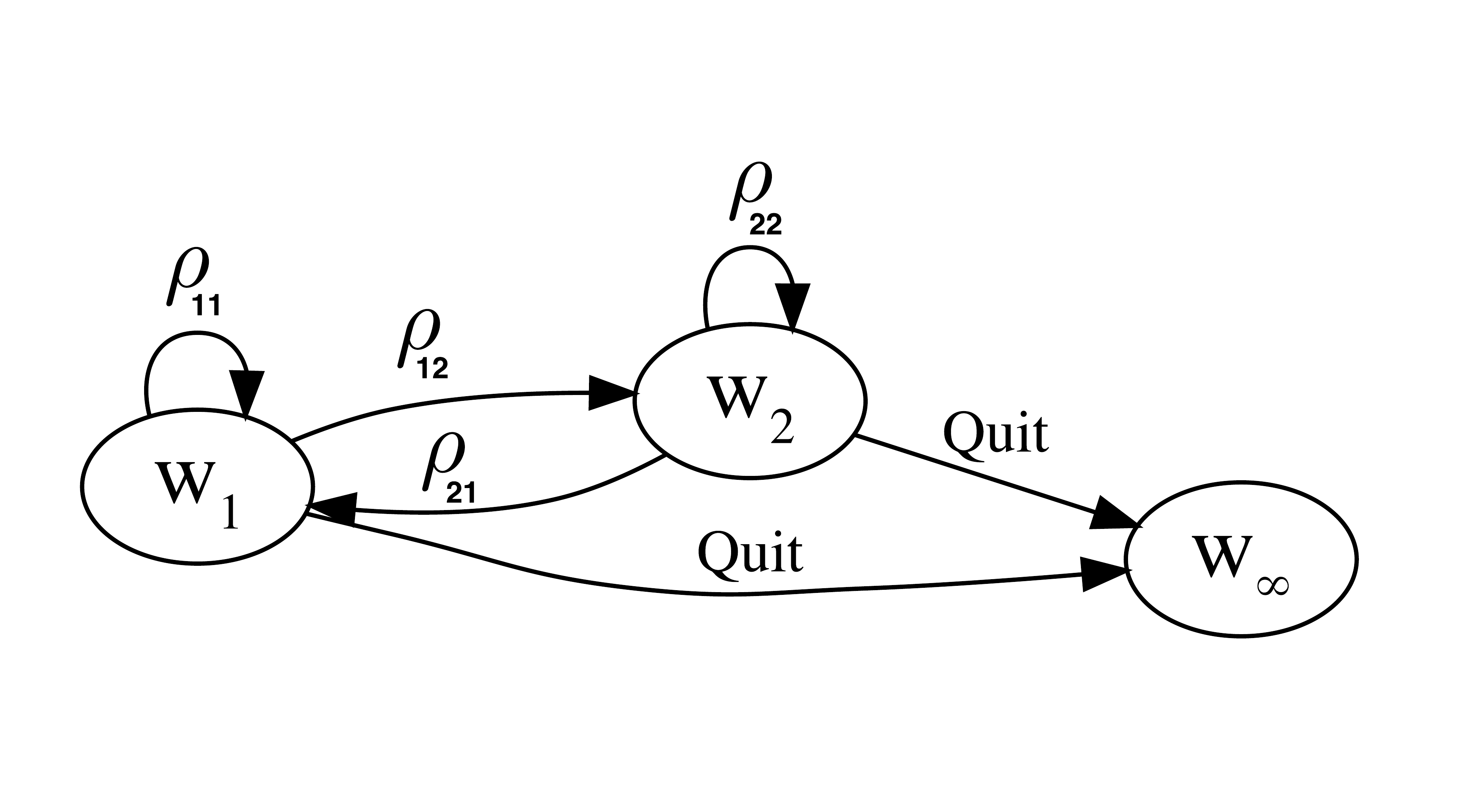}
\caption{SMP representation where $\rho_{ij}$ is the vector of edge labels for the shortest path from $w_i$ to $w_j$ in Figure \ref{fig:smoking_rdceg_a}.}
    \label{fig:semi_markov_smoking}
\end{figure}

In an RDCEG $\calR$, let $V^*$ be the set of vertices from which the edges of $E^*$ emanate and $V'$ be the set of vertices into which these edges enter. The SMP \textbf{Z} for this RDCEG has state space $V \triangleq V^* \cup V'$. Consider a pair $v_i, v_j \in V$ with exactly one edge $e_{ij}$ between them in $\calR$. Let $H_{ij} \sim \textmd{Weibull}(\theta_{ij}, \kappa_{ij})$. Let the parameter $\theta_{ij}$ have an Inverse-Gamma distribution posterior given by $K_{ij} \sim \textmd{IG}(\zeta_{ij}, \beta_{ij})$. Then the compound distribution $G$ has density
\begin{equation}
f_G(t) = \int f_H(t\,|\, \theta, \kappa)f_K (\theta) d\theta.
\label{eq:compound_distribution}
\end{equation}
This gives the holding time distribution on edge $e_{ij}$. The probability of transitioning from $v_i$ to $v_j$ in \textbf{Z} is the same as in $\calR$. If the probabilities of transitioning out of any non-absorbing state do not sum to one in \textbf{Z}, then they can be easily renormalized. In an SMP, there can be at most one edge in a given direction between any two states. See Appendix A of the supplementary material for an SMP representation of an RDCEG which has multiple edges in the same direction between two vertices. Figure \ref{fig:semi_markov_smoking} shows the state transition diagram of an SMP for the smoking cessation RDCEG in Figure \ref{fig:smoking_rdceg_a}.

\textit{Theorem 2.} The SMP representation \textbf{Z} of an RDCEG $\calR$ obtained by the construction described above is a valid semi-Markov process.

If a query concerns only a subset of states $V^\dagger \subset V$, then we can further condense \textbf{Z} to an SMP \textbf{Z}$^\dagger$. See the supplemental document for the proof of Theorem 2 and for the condensed SMP \textbf{Z}$^\dagger$. 

%% file: Sections/Simulation.tex
We next examine the practical performance of our RDCEG model class by studying the evaluation of an intervention designed to reduce falls in the elderly, adapted from \cite{eldridge2005modelling}. To study performance knowing the ground truth, we created artificial datasets but calibrated these to summary statistics provided in various studies to maintain credibility. In the falls intervention, the referral pathways for residents of the community and communal establishments (nursing homes, care homes and hospitals) are different which makes the RDCEG a natural choice of model. An individual's risk of falling is assessed using the Falls Risk Assessment Tool \citep{eldridge2005modelling}. The nature of the treatment provided under the intervention is at the clinician's discretion. So the challenge here lies in assessing the effectiveness of the intervention given the various event pathways an individual might experience.

We simulated 100 instances of open populations of sizes 500, 1500, 2500, 5000, 7500 and 10000 assuming a time-homogeneous setting. Due to time-homogeneity, the entry and exit timing of individuals is not important. Transitions between states are mutinomially distributed and the holding times (see Table \ref{table:times}) are generated from two-parameter Weibull distributions. For each dataset, model selection used the greedy Agglomerative Hierarchical Clustering algorithm across a wide range of prior specifications. Because of the conjugacy properties of this class the search can be evaluated very quickly, see Appendix B of the supplementary material.

\begin{figure}[h!]
\centering
\includegraphics[trim = 0cm 2.5cm 0cm 2.5cm, scale = .35 ]{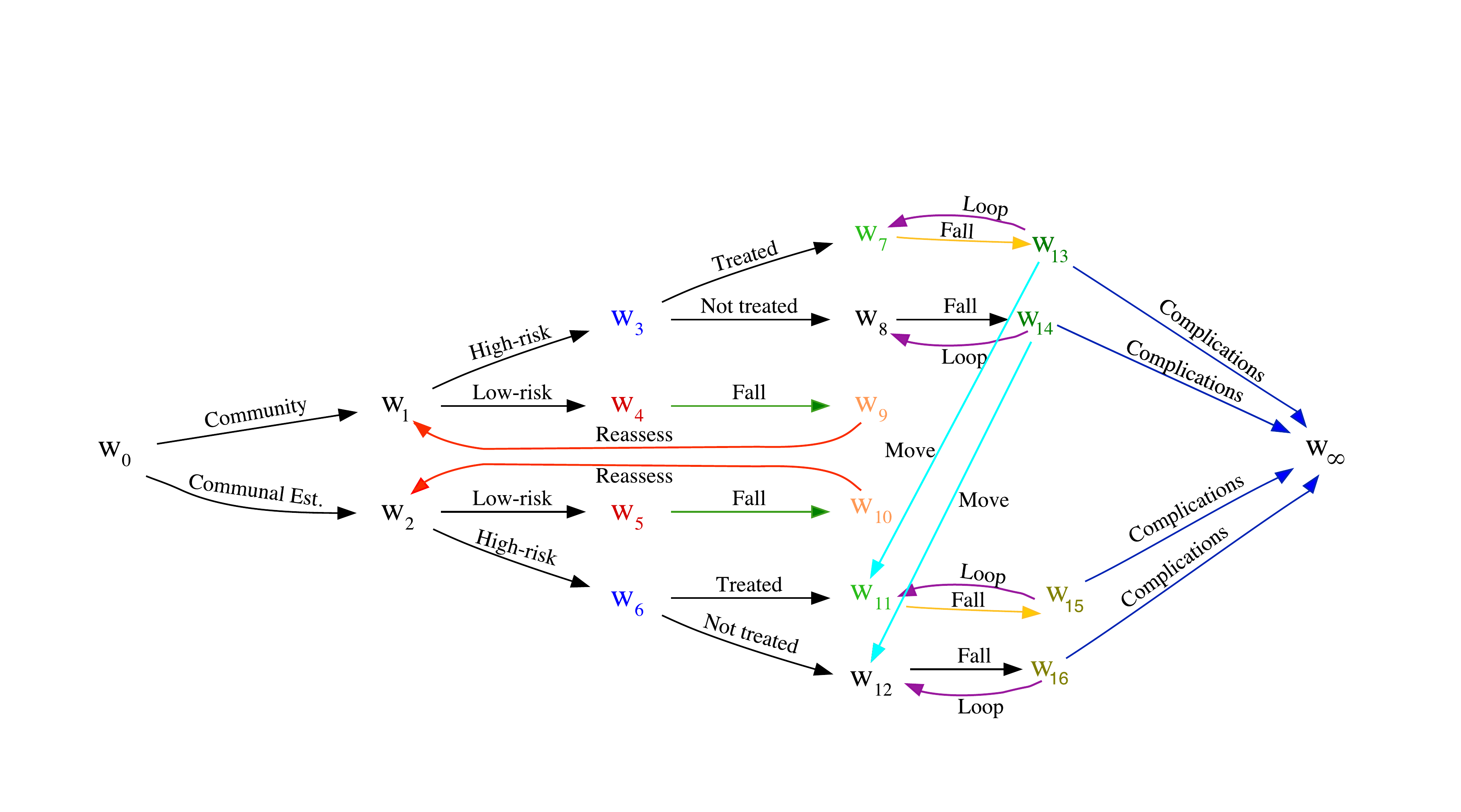}
\caption{The RDCEG model used to generate the falls datasets. Colored vertices represent positions which are in the same stage, and colored edges denote holding time equivalences. Edges from $w_4, w_5, w_7, w_8, w_{11}$ and $w_{12}$ representing no falls have been suppressed.}
    \label{fig:rdceg_falls}
\end{figure}

\begin{table}[h!]
\centering
\resizebox{0.85\columnwidth}{!}{%
\begin{tabular}{|l|l|}
\hline
\textbf{Random variable} & \textbf{Description} \\ \hline
$H_{4,9}, H_{5,10}$ & Duration from assessment to fall for low risk individuals\\
$H_{9,1}, H_{10,2}$ & Duration from fall to reassessment for low risk individuals \\
$H_{7,13},  H_{11,15}$ & Duration from treatment to fall for treated high risk individuals \\
$H_{8,14}, H_{12,16}$ & Duration from assessment to fall for untreated high risk individuals \\
$H_{13, 7}, H_{14, 8}, H_{15, 11}, H_{16, 12}$ & Duration since last fall \\
$H_{13, 11}, H_{14, 12}$ & Duration from fall to moving to a communal establishment \\
$H_{13, \infty}, H_{14, \infty}, H_{15, \infty}, H_{16, \infty}$ & Duration from fall to leaving population due to complications \\
\hline
\end{tabular}%
}
\caption{$H_{i,j}$ refers to the holding time along edge $e_{ij}$ from situations $s_i$ to $s_j$ in Figure \ref{fig:rdceg_falls}.}
\label{table:times}
\end{table}

We were able to retrieve the exact generating model (Figure \ref{fig:rdceg_falls}) for moderately large population sizes (2500 and above) for a wide range of prior specifications. Of course, the accuracy and stability of the average number of stages and clusters across the simulations for varying priors improves as the sample size increases (see Appendix B of the supplement). The model selection algorithm becomes more accurate and more discriminating as it receives more information. As with BNs (see e.g. \cite{silander2007sensitivity}), while several of the MAP models found have minor structural differences (in the number and composition of stages and clusters when compared with the generating model), they are similar in terms of the inference we draw from them. While several measures could be used for this purpose, here we demonstrate this simply by Euclidean and Hellinger distance measures as described below.

Let $\calR$ be the generating RDCEG and $\calR'$ be the model found by the search algorithm. Let $S$ be the set of situations and $E^*$ be the set of edges assigned holding times in the underlying modified tree. The situational error \citep{collazo2016new} for a situation $s \in S$ measures the Euclidean distance between its mean posterior probability $\boldsymbol{\mu}^*(s)$ in $\calR'$ and its true conditional probability $\boldsymbol{\mu}^\dagger(s)$ in $\calR$. The total situation error $\epsilon_\mathbb{U}(\calR, \calR')$ can then be calculated as 
\begin{equation*}
\epsilon_\mathbb{U}(\calR, \calR') = \sum_{s \in S} \norm{\boldsymbol{\mu}^*(s) - \boldsymbol{\mu}^\dagger(s)}_2 .
\end{equation*}

Similarly, define the cluster error for an edge $e \in E^*$ as the Hellinger distance between its posterior density $\text{Wei}(\lambda^*(e), \kappa(e))$ in $\calR'$ and its true holding time density $\text{Wei}(\lambda^\dagger(e), \kappa(e))$ in $\calR$. The total cluster error $\epsilon_\mathbb{C}(\calR, \calR')$ is 
\begin{equation*}
\epsilon_\mathbb{C}(\calR, \calR') = \sum_{e \in E^*} \textmd{d}_{\textmd{H}} (\text{Wei}(\lambda^*(e), \kappa(e)), \text{Wei}(\lambda^\dagger(e), \kappa(e)))
\end{equation*}
\noindent where $\textmd{d}_{\textmd{H}}(f,g) = \sqrt{\tfrac{1}{2} \int (\sqrt{f(x)} - \sqrt{g(x)})^2 dx}$ denotes the Hellinger distance between two densities $f$ and $g$, and $\lambda^*(e)$ is the posterior mean of the Weibull-Inverse-Gamma compound distribution in $\calR'$ (see Appendix A of the supplementary material). 

\begin{figure}[h!]
\begin{subfigure}{0.30\textwidth}
\centering
  \includegraphics[trim = 0cm 0cm 0cm 0cm, scale = .26 ]{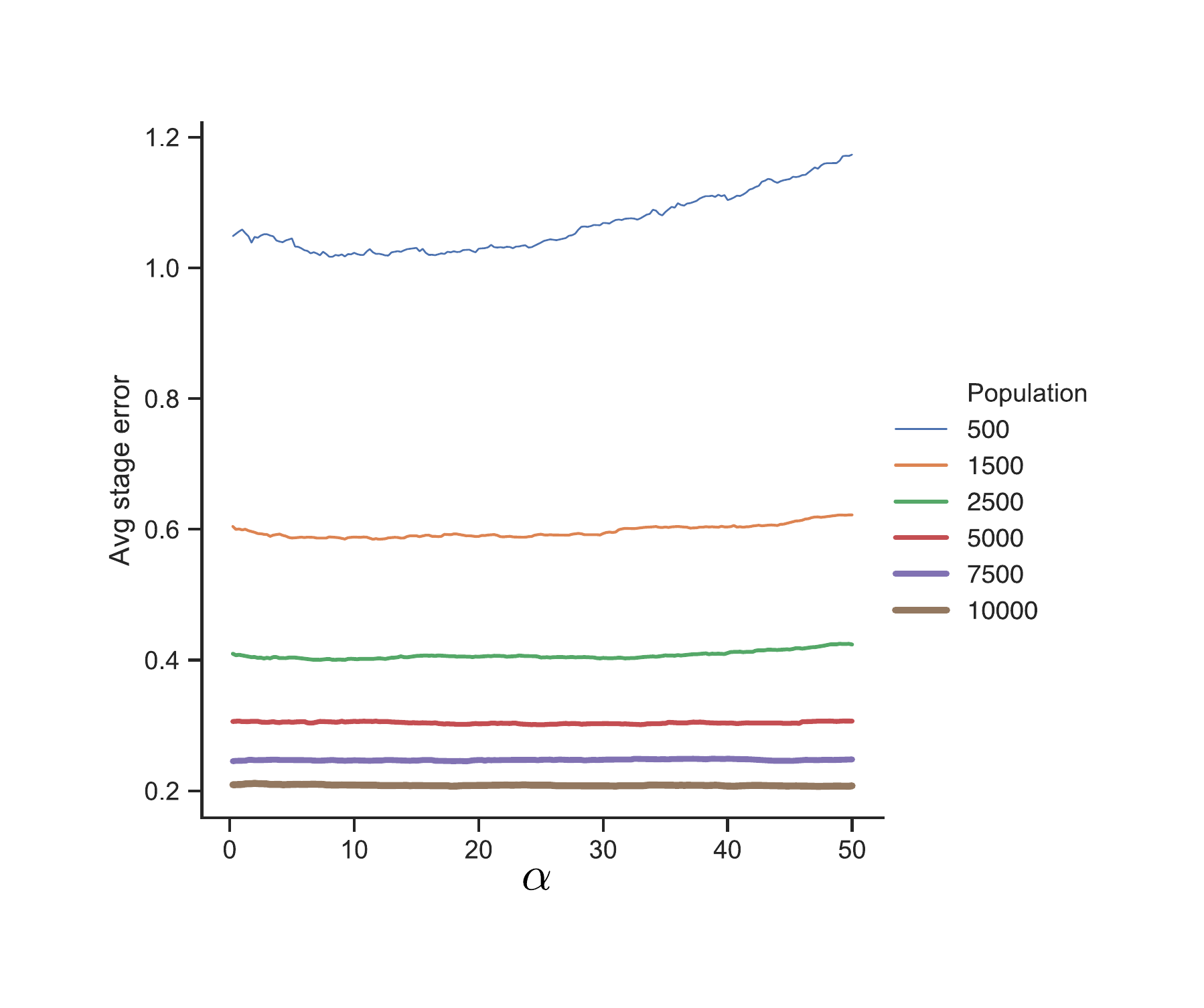}
  \caption{Average situational error}
    \label{fig:avg_error}
\end{subfigure}
\begin{subfigure}{0.68\textwidth}
\centering
  \includegraphics[trim = 0cm 0cm 0cm 0cm, scale = .31 ]{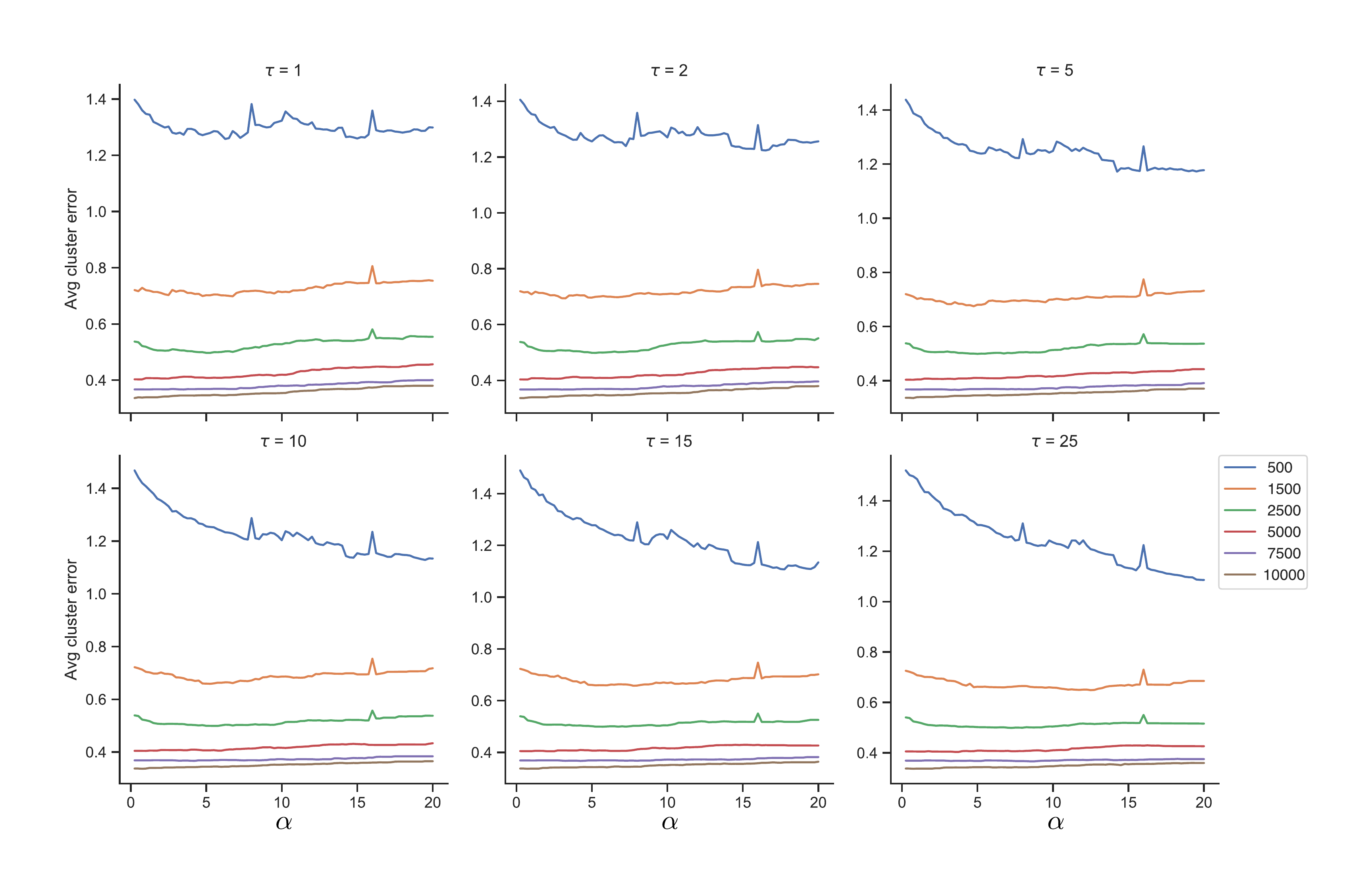}
  \caption{Average cluster error}
    \label{fig:avg_cluster_error}
\end{subfigure}
\caption{Averages taken over 100 simulations for each sample size. $\tau$ denotes the phantom holding time and $\alpha$ is the total number of phantom units.}
\end{figure}

From Figures \ref{fig:avg_error} and \ref{fig:avg_cluster_error}, we can see that the probabilistic accuracy of the models improves very quickly in response to moderate increases in the population size. The total situational and cluster errors are relatively higher and more subject to volatility for the population size of 500. However, for population sizes of 1500 and more, there is increased stability across prior specifications with the situational and cluster errors consistently lower than 0.5 and 0.8 respectively.

Some of the higher scoring competing models retrieved by the search algorithm are given in the supplementary material. During the search process, some high scoring models identified $\{w_{1}, w_{2}\}$, $\{w_{8}, w_{12}\}$ and $\{w_{7}, w_{8}, w_{11}, w_{12}\}$ as stages. These models failed to identify that the high-risk to low-risk proportion, and the risk of falling among non-treated individuals are different for the community and communal establishment. Of these, the misidentification of relatively most significant consequence is $\{w_{7}, w_{8}, w_{11}, w_{12}\}$ indicating that treatment does not reduce the probability of falling. However, this stage was only selected by the algorithm for a particular set of priors and for the small population size of 500. In clusters, misclassification occurred for edges $\{e_{13,11}, e_{14,12}, e_{12,16}\}$ which is of little significance. 

In heterogeneous populations, information is often skewed against the smaller vulnerable groups. Here, we found that there is disparity between the observations of high-risk and low-risk individuals. In spite of this, the RDCEG framework performs well for both risk groups. Note that in this analysis the RDCEG model class used assumes recurrent falls to be mutually independent. Experts might argue that an abnormally large number of falls could be indicative of the individual suffering from a chronic condition like Parkinsonism. In the RDCEG framework it is easy to embellish the model to incorporate such new hypotheses simply by adding new well-defined positions. Indeed such embellishments using expert judgments would be encouraged through discussion of the graphical interface. 

%% file: Sections/Case.tex
\begin{figure}[h!]
\centering
  \includegraphics[trim = 10cm 0.5cm 11.5cm 0cm, width = .50\textwidth]{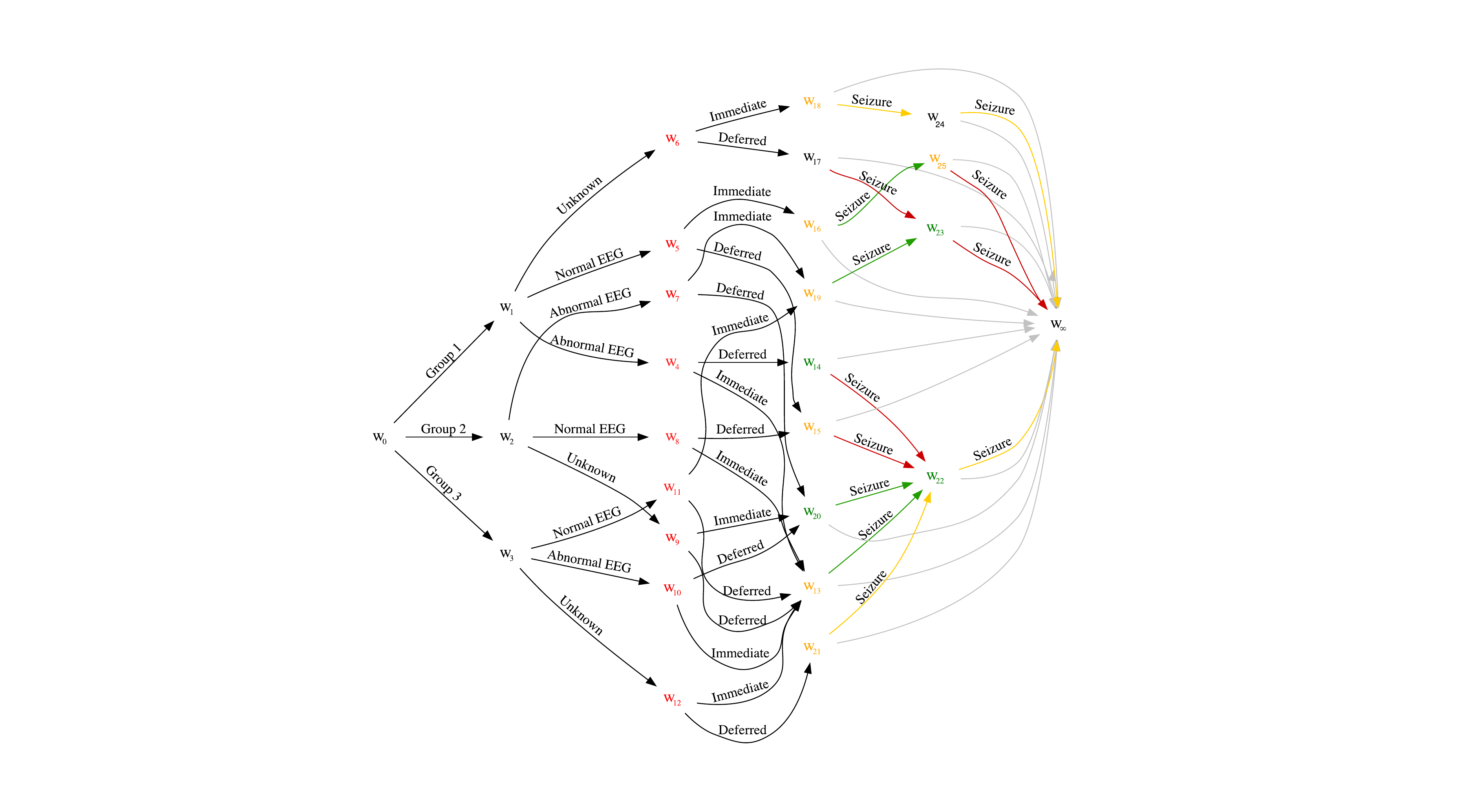}
\caption{The MAP simplified RDCEG for the epilepsy case study. The grayed out edges are labeled ``No more seizures".}
    \label{fig:rdceg_mess}
\end{figure}

We next investigate the effects of antiepileptic drug treatment in individuals presenting with single seizures or symptoms of early epilepsy. Generally the benefits outweigh the risk of side-effects (teratogenicity, drug and dose related toxicity etc.) for those diagnosed with epilepsy. However, the balance of benefits and risks is not as clear for those where the diagnosis is uncertain. Our data was collected in the 1990s as part of the MRC Multicentre trial for Early Epilepsy and Single Seizures \citep{marson2005immediate} in which individuals with early epilepsy and single seizures were randomized to immediate or deferred treatment. For the sake of brevity, we focus our analysis on a subset of this data. In this article, we examine the effects of age and EEG abnormality on the occurrence of the first and second seizure after randomization in the two treatment groups (total = 1420). Note that this simple setting has only two passage-slices and so we use a simplified RDCEG with no cyclic edges. Based on expert recommendations, we classified ``non-specific" type of EEG abnormality as a normal EEG. We created three age categories: Group 1 (0-20 years), Group 2 (20-40 years) and Group 3 (40 $\geq$ years). The priors were informed by data in the existing domain literature. The RDCEG fitted to the data is shown in Figure \ref{fig:rdceg_mess}. Code for data-cleaning and model selection is provided within the supplementary material. 

The graph in Figure \ref{fig:rdceg_mess} summarizes the MAP model within the RDCEG class for our population of interest. It comprises people who present with symptoms of early epilepsy and are prescribed anti-epileptic drugs. All the conclusions here are made for individuals who remain in the population. The graph shown in Figure \ref{fig:rdceg_mess} drew out a new feature not previously noticed by domain experts. We see that the outcomes in terms of the probability of having the first and second seizure after randomization appear to be independent of the individual's age group given that (s)he has an abnormal EEG but not otherwise (from the stages of the cut $\{w_i: i = 4, 7, 10\}$ in the subgraph induced by individuals with abnormal EEGs). For example, \cite{kim2006prediction} reported that this independence relationship was applicable to the entire population. Additionally, the MAP model appears to suggest that the three age groups have different compositions of normal, abnormal and unknown EEG types. All individuals who have a normal EEG and those with an abnormal EEG immediately prescribed to drugs have a probability of 49.50\% of having the first seizure. Among these individuals, those who have had a seizure have a 66.40\% of having a second seizure. Individuals with an abnormal EEG who have been assigned to a deferred treatment group have a high risk of 66.40\% for the first and second seizure. This suggests that the deferral of treatment might negatively impact individuals with an abnormal EEG. 

We also discovered some interesting holding time patterns in our model. For most individuals with known EEG results, the holding time for the first seizure (mean 359.98 or 593.53 days) is longer than that for the second seizure (mean 186.21 days from the first seizure). However, our model suggests that individuals on paths given by the edge labels \{Group 1, Normal EEG, Immediate\}, \{Group 2, Abnormal EEG, Immediate\} and \{Group 3, Normal EEG, Immediate\} have a longer holding time for the second seizure (mean 583.53 days) than the first seizure (mean 359.98 days). Such discoveries were fed back to domain experts to examine their plausibility and their consequences. The RDCEG graph is thus a vehicle through which such creative interchanges can be separated so that models can be further enhanced. Further analyses of this example can be found in Appendix C of the supplementary material.

%% file: Sections/Discussion.tex
Here we have demonstrated the efficacy of the RDCEG model for use with certain types of heterogeneous public health datasets. There are various extensions that could now be applied. As we are modeling open populations, in some circumstances it appears necessary to model \textit{how} individuals enter the target population. For example, there may be different referral pathways for different sections of the society. It would also be interesting to study how inference from an RDCEG can be extended across different populations by identifying which structural aspects of its graph are associated with institutional developments and which are associated with a population's physiology. 

Extensions of the methodology to incorporate additional heterogeneities introduced by interdependence of the adverse events would be beneficial for the domains of reliability, survival, ecological and conservative studies, and are currently being explored. The motivation for such developments come from conditional frailty models \citep{box2006repeated}. We are also currently developing algorithms that can draw out all conditional independence statements and their corresponding expressions in natural language within a given RDCEG in analogy to the d-separation theorem of \cite{pearl2009causality}.

Currently, Bayesian hierarchical models which piece together noisy information available from varied sources in a structured and formally defensible manner are being developed with the RDCEG playing a central role, see  \cite{bunninbayesian2019}. So far these have only been developed for simple models with no model selection applied. A further next step here would be to consider a more fluid RDCEG with a non-stationary staging and clustering structure across passage-slices akin to non-stationary Dynamic Bayesian Networks \citep{grzegorczyk2009non, song2009time}. This involves methods based on Bayesian multiple change-point processes and regularized auto-regression. However, although there is much still to be explored, we believe that the framework described above provide very promising additional tools for the public health analyses of open populations.

%% file: Appendices/AppendixA.tex
\subsection{Probability Measure on an Event Tree}
\label{sec:probability_measure}

Consider an event tree $\calT = (V, E)$. For a vertex $v \in V$ with set of children given by $ch(v)$ define a \textit{floret} as $\calF(v) = (ch(v), E(v))$ where $E(v) = \{e_{vv'} : v' \in ch(v)\}$. For two event trees $\calT_1, \calT_2$ rooted at vertex $v_0$ write $\calT_1 \preceq \calT_2$ if $\calT_1$ is a subtree of $\calT_2$. Say $\calT_1$ is a \textit{minimal coarsening} of $\calT_2$ if $\calT_1$ can be constructed from $\calT_2$ by deleting exactly one floret. An event tree of an infinite size can thus be constructed from a finite sized tree by sequentially adding florets. 

Let event tree $\calT \triangleq (\calT_1, \calT_2, \ldots)$, where $\calT_j \preceq \calT_{j+1}$, each $\calT_j$ is finite and $\calT_j$ is a minimal coarsening of $\calT_{j+1}$, $j \in \mathbb{N}$. For each tree $\calT_j$, $j \in \mathbb{N}$ we can define a probability space given by $(\Omega_j, \mathscr{F}_j, \mathbb{P}_j)$. For any tree $\calT'$, let $\Lambda(\calT')$ be the set of all root-to-leaf paths in $\calT'$. Every atom $\lambda \in \Lambda(\calT')$ can be thought of as an outcome trajectory for a unit passing through the tree. Thus $\Omega_j$ for $\calT_j$ corresponds to $\Lambda(\calT_j)$ and $\mathscr{F}_j$ is the set of all possible unions of $\lambda \in \Lambda(\calT_j)$. 

The sample space $\Omega$ of $\calT$ can then be written as an infinite product space given by $\Omega \coloneqq \Omega_i \times \Omega_2 \times \ldots$, which has a product $\sigma\textmd{-algebra}$ $\mathscr{F}$. As this product is countable, the $\sigma\textmd{-algebra}$ $\mathscr{F}$ is generated by cylinder sets given by
\begin{align*}
C = \{(\omega_1, \omega_2, \ldots) \in \Omega : \omega_{1} \in A_{1}, \omega_{2} \in A_{2}, \ldots, \omega_{k} \in A_{k}\},    
\end{align*}
for some $k \in \mathbb{N}$ and $A_{i} \subseteq \mathscr{F}_{i}$ for $1 \leq i \leq k$. Each cylinder set describes an outcome trajectory for a unit from the root up to $k$ transitions. The subsequent evolution of the outcome trajectory can be arbitrary. Each outcome trajectory then has an associated probability (trivially each root-to-leaf path of this finite tree can be assigned an equal probability). The probability of the cylinder set $C$ is then given by
\begin{align*}
    \mathbb{P}_C = \mathbb{P}_{1}(A_{1}) \times \ldots \times \mathbb{P}_{k}(A_{k}).
\end{align*}
\noindent Note that the collection of finite unions of these cylinder sets forms an algebra $\mathscr{F}_0$. Thus $\mathscr{F}$ is the  $\sigma\textmd{-algebra}$ generated by $\mathscr{F}_0$.  

Given the probability spaces $(\Omega_j, \mathscr{F}_j, \mathbb{P}_j)$, $j \in \mathbb{N}$, for a finite subset $I$ of $\mathbb{N}$, let the product measure on $\Omega_I$ be denoted by $\mathbb{P}_I$. Since all $(\Omega_j, \mathscr{F}_j)$, $j \in \mathbb{N}$ are Borel spaces and the sequence of probability measures $\mathbb{P}_I$, $I \subset \mathbb{N}$ is a consistent family of finite-dimensional distributions by the construction given above, by Kolmogorov's Extension theorem there exists a unique probability measure $\mathbb{P}$ on the infinite product space $\Omega$ that agrees with the measures $\mathbb{P}_I$ on $\Omega_I$, $I \subset \mathbb{N}$.  
\subsection{An RDCEG as an Infinite Chain Event Graph}

\begin{figure}[!ht]
\centering
\begin{subfigure}{0.30\textwidth}
  \centering
  \includegraphics[trim = 0cm 7cm 0cm 2.5cm, scale = .12 ]{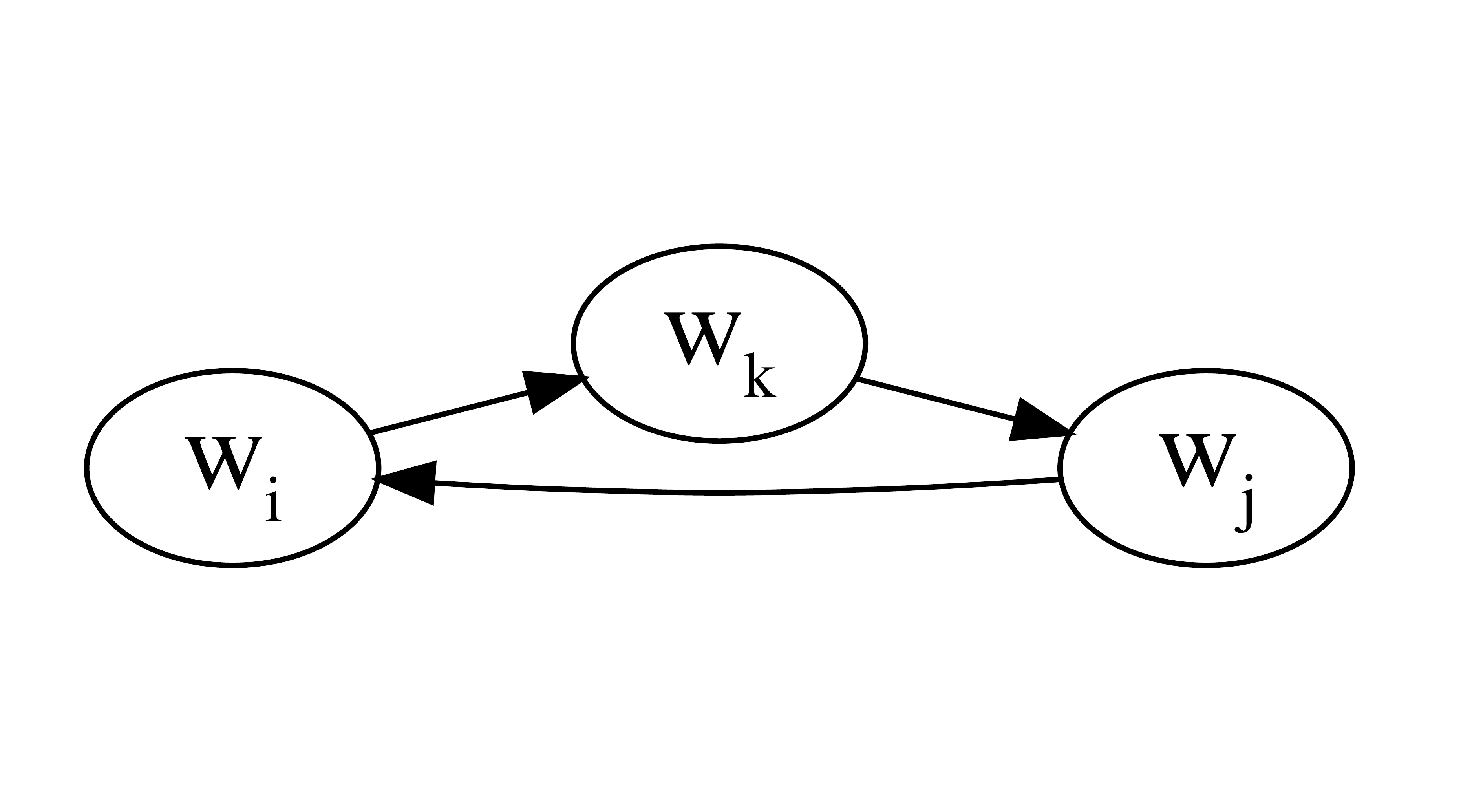}
  \caption{}
   \label{fig:rolled_example_a}
\end{subfigure}
\begin{subfigure}{0.69\textwidth}
\centering
  \includegraphics[trim = 0cm 5cm 0cm 2.5cm, scale = .25 ]{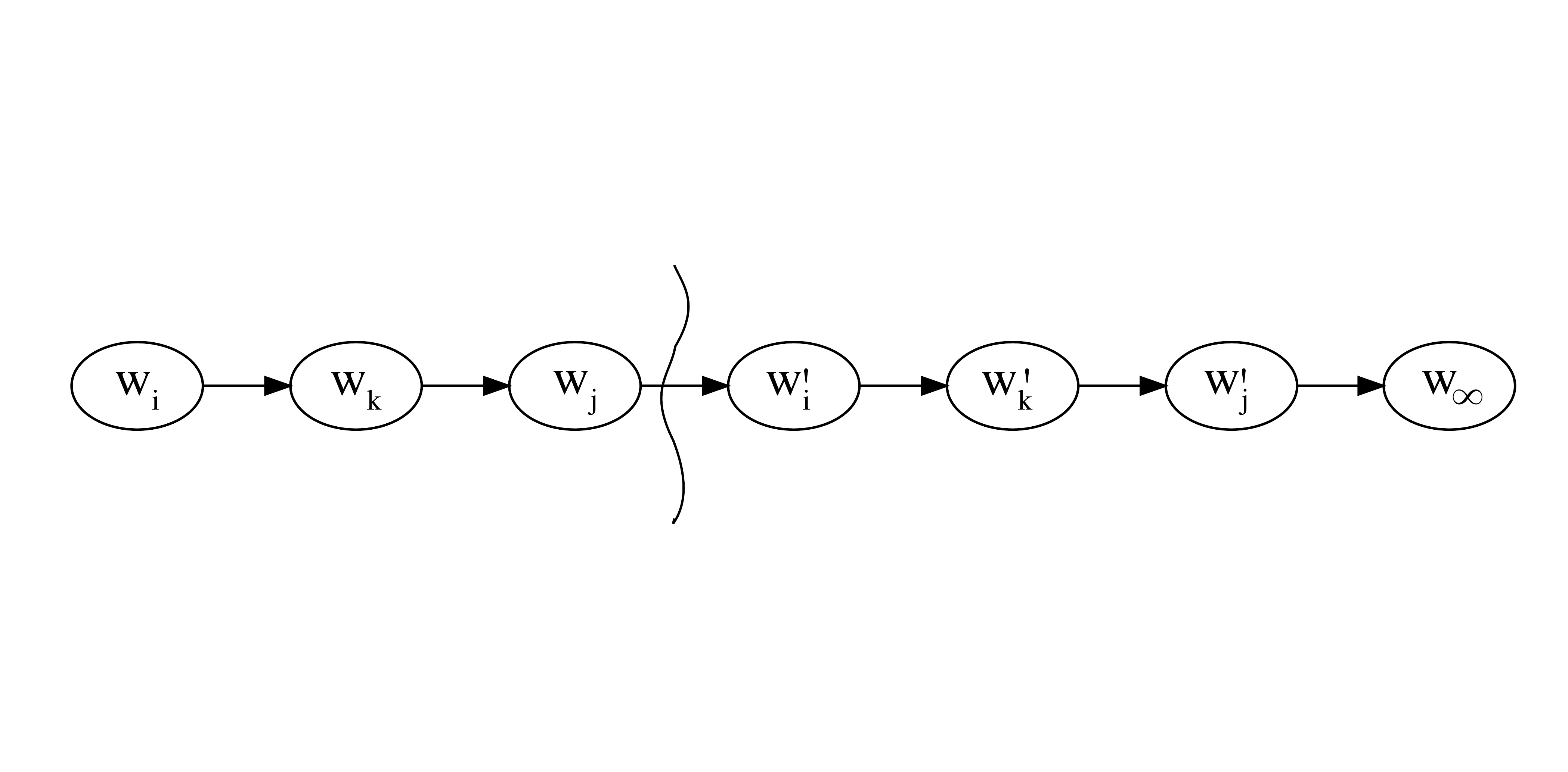}
  \caption{}
  \label{fig:rolled_example_b}
\end{subfigure}
\caption{(a) An RDCEG $\calR$; (b) $\calC_{2}$ of $\calR$. The wavy line separates $\calP_1$ on the left from $\calP_2$ on the right. The leaf $w_j'$ of $\calP_2$ is collected into the sink $w_\infty$.}
\label{fig:rolled_example}
\end{figure}

Consider an RDCEG $\calR$. Let its infinite CEG rolled out up to $n$ passage-slices be denoted by $\calC_{n}$. The construction of the graph of $\calC_n$ begins with the first passage-slice $\calP_1$ of the RDCEG $\calR$. The cyclic edges of $\calR$ then connect the leaves of $\calP_1$ to the subgraphs of $\calR$ representing its second passage-slice $\calP_2$. The remaining passage-slices $\calP_k$, $k = 3,\ldots,n$ are iteratively added to $\calC_n$ following the procedure described above. Thus the cyclic edges of the RDCEG connect $\calP_j$ to $\calP_{j+1}$, $j = 1,\ldots, n-1$. The infinite CEG $\calC_n$ has one sink vertex $w_\infty$ and it collects all the critical terminating paths of the RDCEG $\calR$. Additionally, all the leaves of $\calP_n$ are also collected into the sink $w_\infty$. For a simple illustration, see Figure \ref{fig:rolled_example}.

\subsection{Proof of Theorem 1}

Consider an RDCEG $\calR$. We can roll out $\calR$ into an infinite CEG as described above to query dependence relationships concerning events stretching across passage-slices. For all practical queries, we find it sufficient to consider an infinite CEG up to some finite depth of $n$ passage-slices, denoted by $\calC_n$, $n \in \mathbb{N}$. To prove the statement of Theorem 1, we show that by expressing an RDCEG as an infinite CEG, standard CEG technologies can be applied to read conditional independence statements directly from the graph topology. 

For an RDCEG $\calR$, consider a rolled out infinite CEG $\calC_n = (V, E)$, $n \in \mathbb{N}$. Let $\Lambda(\calC_n)$ be the set of all root-to-sink paths in $\calC_n$. These paths are the atoms generating the sigma algebra of events of the underlying probability space $\Omega$. In this graph, we can query events which are \textit{intrinsic}. An event $\mathcal{E}$ - which is the union of a set of atoms say $\Lambda_\mathcal{E}$ - is said to be intrinsic to $\calC_n$ if and only if the subgraph $\calC_n(\Lambda_\mathcal{E})$ induced by these atoms has exactly those root-to-sink paths which constitute $\Lambda_\mathcal{E}$. That is, the map $\Lambda_\mathcal{E} \mapsto \calC_n(\Lambda_\mathcal{E})$ is invertible \citep{collazo2018chain, thwaites2015separation}. For instance, in Figure \ref{fig:rolled_out_ceg} in the article, consider the ordered sets of vertices $\{w_0, w_1, w_0', w_1', w_\infty\}$ and $\{w_0, w_2, w_0', w_2', w_\infty\}$ representing paths A and B respectively. Both paths are themselves intrinsic. However, the event given by the union of paths A and B is not intrinsic as the subgraph induced by this event also includes paths on $\{w_0, w_1, w_0', w_2', w_\infty\}$ and $\{w_0, w_2, w_0', w_1', w_\infty\}$. 

For each vertex $v \in V \backslash\{w_\infty\}$, we can define \textit{incident variables} as follows

\begin{equation}
I(v) = 
\begin{cases}
1, & \{\lambda: v \in \lambda\} \\
0, & \{\lambda: v \notin \lambda\}
\end{cases}
\quad \text{and} \quad
J(v) = 
\begin{cases}
v', & \{\lambda: (v, v') \in \lambda\} \\
0, & \{\lambda: (v, v') \notin \lambda\}
\end{cases}
\label{eq:const}
\end{equation}

Variable $I(v)$ indicates whether a path passes through vertex $v$. Variable $J(v)$ denotes which position is occupied next by an individual at $v$. These random variables are measurable with respect to the set $\Lambda(\calC_n)$ of $\calC_n$ and its associated probability $\Omega$ . We now define a variable $X(v)$ as $J(v) | I(v) = 1$ for $v \in V \backslash \{w_\infty\}$. This variable tells us where an individual who has arrived at a vertex $v$ transitions to after arriving there. 

Consider a set of vertices representing a cut $U$ in $\calC_n$. Denote by $\textbf{P}(U)$ the vector of variables $\{X(u): u \in U\}$. Let $Pa(U)$ be the set of parents of the vertices in $U$. Denote by $\textbf{Q}(U)$ the vector of variables $\{X(u^*): u^* \in Pa(U)\}$. Let $\Lambda_v$ be the set of paths in $\calC_n$ from the root to a vertex $v$. Denote by $\textbf{R}(U)$ the vector of variables $\{X(u^{**}): u^{**} \in \lambda_u, \lambda_u \in \Lambda_u\}$ with state space $\lambda_u$ representing the vertices visited by an individual before arriving at some $u \in U$. 

The associated probability mass functions for each of these random variables are
\begin{equation*}
p_{\textbf{P}(U)}(u') = \pi_{uu'}, \quad p_{\textbf{Q}(U)}(u) = \textstyle \sum_{\lambda_u \in \Lambda_u} \textstyle \prod_{v \in \lambda_u, v \notin U} \pi_{vv'}, \quad p_{\textbf{R}(U)}(\lambda_u) = \textstyle \prod_{v \in \lambda_u, v \notin U} \pi_{vv'}
\end{equation*}
\noindent where $\pi_{vv'}$ is the probability of transitioning to vertex $v'$ given that the individual is at vertex $v$.

By construction of these variables, we get that 
\begin{equation}
\textbf{R}(U) \Perp \textbf{P}(U) \,|\, \textbf{Q}(U)
\label{eq:conditional}
\end{equation}
\noindent where $\Perp$ indicates probabilistic independence. This statement indicates that once we know that an individual has arrived at stage $u$, where (s)he goes immediately after leaving $u$ is independent of his/her path into $u$. For a fine cut $W$, we can define $\textbf{P}(W)$ to be a vector of variables representing the possible future paths from some $w \in W$ up to a finite depth. Then Equation \ref{eq:conditional} could be interpreted as follows. Given that the individual has arrived at position $w$, where (s)he goes after leaving $w$ - up to the finite depth (in terms of passage-slices) chosen - and how long he spends in these states is independent of his/her path into $w$. Since the graph (or subgraph) of an RDCEG with cyclic edges repeats when it is drawn out as an infinite Chain Event Graph, these statements can be read directly from the cuts and fine cuts of the RDCEG graph itself. This is described by the statement of Theorem 1. We can read conditional independence statements from subgraphs of the RDCEG or for specific events as long as they are intrinsic.

This result follows from the definition of a stage and position (see Section \ref{sec:rdceg} of the article). Observing vertex $w$ in a fine cut for an individual in Theorem 1 corresponds to observing the position $w$ (s)he occupies. By the definition of a position, this is sufficient to determine his/her complete probabilistic future, or equivalently, his/her future up to $n$ passage-slices. Thus any additional information about how (s)he got to $w$ is superfluous. Similarly, observing the stage $u$ in a cut occupied by an individual enables us to probabilistically determine his/her next step in the evolution of the process.

We do not explicitly depict dropouts which are not critical terminating events (see Section \ref{sec:rdceg} of the article) in the graph of an RDCEG. If we were to depict these dropouts and collect them in a separate sink node $d$, we could define a fourth variable for $v \in V \backslash \{w_\infty, d\}$
\begin{equation*}
K(v, t) = 
\begin{cases}
1, & \{\lambda: J(v) \neq d \,\,\textmd{before or at transition}\,\, t\} \\
0, & \{\lambda: J(v) = d \,\, \textmd{before or at transition}\,\, t\}.
\end{cases}
\end{equation*}
\noindent This would be equivalent to the construction in \ref{eq:const} by redefining $X(v)$ as $X(v, t)$ given by $J(v) | \{I(v) = 1, K(v, t) = 1\}$. Here, $X(v, t)$ is a random variable indicating the next vertex occupied by an individual who has arrived at vertex $v$ and has not entered a dropout state $d \in L(\calT) \backslash D^*$ (where $\calT$ is the underlying event tree) in the next $t$ transitions. In this way, for a cut $U$, we would be interested in the variable $X(v, 1)$ when defining $\textbf{P}(U)$. Similarly, to enable a discussion about the conditional independence relationships in the next $n$ passage-slices in a fine cut $W$, the variable of interest when defining $\textbf{P}(W)$ would be $X(v, k)$ where $k$ is the maximum number of transitions an individual could make from a vertex in $W$ up to the next $n$ passage-slices.

\textit{Example 1.} We now show how the conditional independence statement of Example 1 in the article can be read through the infinite CEG representation. Consider the subgraph of $C_2$ showed by the darkened edges in Figure \ref{fig:rolled_out_ceg} in the article. The collection $W = \{w_1, w_2\}$ is a fine cut. $\textbf{Q}(W)$ indicates whether the man used the cessation services in his last attempt. We have $\textbf{R}(W) = \{X_{w_0}\}$ and $\textbf{P}(W) = \{X_{w_0'}, X_{w_1'}, X_{w_2'}\}$. From cut $W$, we can say that so long as the man does not drop out before his next attempt, his choice of availing the services in that attempt and the success of that attempt are not dependent on whether he availed these services in his previous attempt(s). 

\subsection{SMP Representation for Multiple Edges Between Two Vertices}
\label{subsec:smp_rep}

Consider an RDCEG $\calR$ for which $V^*$ is the set of vertices from which the edges of $E^*$ (see Section \ref{sec:rdceg} of the article) emanate and $V'$ is the set of vertices into which these edges enter. The state space of \textbf{Z} is $V \triangleq V^* \cup V'$. Suppose there are $m$ edges $e_{ij}^l$, $l = 1,\ldots,m$ from $v_i$ to $v_j$, for $v_i, v_j \in V$. For a valid SMP representation, edges $e_{ij}^l$, $l = 1,\ldots,m$ need to be coalesced into a single edge in \textbf{Z}. Without loss of generality, assume that the elements of vector $(\pi_{i1},\ldots, \pi_{im})$ represent the probability of going along edges $e_{ij}^1, \ldots,e_{ij}^m$ respectively given that the individual has arrived at vertex $v_i$ in $\calR$ and does not dropout at the next transition. 

Let $G^l$ be the compound distribution given by Equation \ref{eq:compound_distribution} in the article for each edge $e_{ij}^l$. Then the holding time distribution $G$ on edge $e_{ij}$ from state $v_i$ to state $v_j$ in \textbf{Z} is a mixture distribution with density
\begin{equation}
f_G(t) = \sum_{l = 1}^m \pi_{il}\, f_{G^l}(t),
\label{eq:mixture_distribution}
\end{equation}
where $\pi_{il}$ acts as the weight on edge $e_{ij}^l$ in the mixture distribution, $l = 1,\ldots,m$ and the transition probability from state $v_i$ to state $v_j$ in \textbf{Z} is given by $\pi(v_i, v_j) = \sum_{l = 1}^m \pi_{il}$.

\subsection{Proof of Theorem 2}
\label{sec:proof_theorem_semi}

By construction, the representation \textbf{Z} of the RDCEG $\calR$ has state space \textbf{S} = $V^* \cup V'$ where $V^*$ and $V'$ are defined as above. Recall that an SMP is completely defined by its initial distribution and its renewal kernel.

Let $v_i, v_j \in$ \textbf{S}. The transition matrix \textbf{P} = $(p_{ij})$ of \textbf{Z} is given by entries
\begin{equation*}
p_{ij} =
\begin{cases}
\pi_{ij}^*, & \textmd{if}\, \mathbbm{1}_{\calR}(v_i \rightarrow v_j) = 1\\
0, & \textmd{otherwise}
\end{cases}
\label{eq:transition_matrix}
\end{equation*}
\noindent where $\mathbbm{1}_{\calR}(v_i \rightarrow v_j)$ indicates whether there exists at least one edge from $v_i$ to $v_j$ in the RDCEG $\calR$ and $\pi_{ij}^* = \sum_{l = 1}^{m} \pi_{ij}^l$ where $\pi_{ij}^l$ is the transition probability associated with edge $e_{ij}^l$, $l = 1,\ldots,m$ for the $m$ edges from $v_i$ to $v_j$. If all the $k$ edges emanating from $v_i$ in the RDCEG $\calR$ belong to $E^*$, then $\sum_{n=1}^{k} p_{in} = 1$. In some cases, we may have $\sum_{n = 1}^{k} p_{in} < 1$ where some edges emanating from vertex $v_i$ in $\calR$ belong to $E \backslash E^*$. In such cases, a standard holding time can be chosen to be set on such edges or the probabilities on the remaining edges can be renormalized. 

The cumulative conditional holding time distribution at edge $e_{ij}$ from state $v_i$ to state $v_j$ in \textbf{Z} is
\begin{equation*}
F_{ij}(t) = \mathbb{P}(\tau_{ij} \leq t \,|\, X_{n+1} = v_j, X_{n} = v_i),
\label{eq:holding_time_Z}
\end{equation*}
\noindent where the holding time $\tau_{ij}$ in \textbf{Z} is governed by the distribution in Equation \ref{eq:compound_distribution} in the article or by Equation \ref{eq:mixture_distribution} in Appendix A.

The entries of the renewal kernel \textbf{Q} of the representation \textbf{Z} are given by
\begin{equation*}
Q_{ij}(t) = p_{ij}F_{ij}(t),
\label{eq:renewal_kernel_Z}
\end{equation*}
\noindent and the initial distribution vector $\textbf{p}$ has entry $1$ for the vertex of \textbf{Z} where the individual enters the system and $0$ elsewhere. The renewal kernel \textbf{Q} and initial distribution $\textbf{p}$ completely describe the stochastic process \textbf{Z}. 

\subsection{The Condensed SMP}

Using the notation from Section \ref{sec:semimarkov} of the article, if a query concerns only a subset of states $V^\dagger \subset V$, then the query can be interrogated using the SMP \textbf{Z}$^\dagger$. Consider $v_i, v_j \in V^\dagger$. If there exists an edge or multiple edges between $v_i$ and $v_j$, then the conditional transition probability and holding time distributions are derived as described in Section \ref{sec:semimarkov} of the article and Section \ref{subsec:smp_rep} of Appendix A.

Suppose there is no edge between $v_i$ and $v_j$ in the RDCEG $\calR$ but there exists a path $\epsilon_{ij}$ from $v_i$ to $v_j$ in the passage-slice $\calP_1$ such that the vertices between $v_i$ and $v_j$ on the path are not in $V^\dagger$. This path is then condensed into an edge $e_{ij}$ from $v_i$ to $v_j$ in \textbf{Z}$^\dagger$. Let the holding time distribution along the $l$th edge in $\epsilon_{ij}$ be governed by the compound distribution $G^l$ and the conditional transition probability distribution be given by $\pi^l$. As the holding time distributions are assumed \textit{a priori} independent, the holding time distribution $G$ on the edge $e_{ij}$ in \textbf{Z}$^\dagger$ is given by the convolution
\begin{equation*}
f_G(t) = (f_{G^1} * \ldots * f_{G^l})(t),
\label{eq:convolution_distribution}
\end{equation*}
\noindent where $*$ represents a convolution. The conditional transition probability from state $v_i$ to state $v_j$ in \textbf{Z}$^\dagger$ is given by
\begin{equation*}
\pi_{ij} = \prod_{l = 1}^m \pi^l,
\label{eq:convolution_transition}
\end{equation*}
\noindent where $m$ is the number of edges in $\epsilon_{ij}$ and the conditional transition probabilities are \textit{a priori} independent. If the probability of transitioning out of a state does not sum to one, the probabilities on its emanating edges should be renormalized. If the density function or mean of a compound distribution cannot be calculated in closed-form, then this can be done by applying Monte Carlo integration methods. Note that in most cases convolutions of probability distributions cannot be solved analytically and will have to be handled numerically. For details on the convolutions of Inverse-Gamma distributions see \cite{witkovsky2001computing, giron2001note}. 

\subsection{Mean and Variance of the Weibull - Inverse Gamma Compound Distribution}

Suppose variable $X$ is distributed according to a Weibull distribution $G$ with scale parameter $\theta$ and shape parameter $\kappa$. The scale parameter $\theta$ comes from an Inverse-Gamma distribution $F$ with shape parameter $\zeta$ and scale parameter $\beta$. The resulting compound holding time distribution $H$ is obtained by marginalizing $G$ over $F$ as follows
\begin{equation*}
    f_{H}(x) = \int f_G(x|\theta, \kappa) f_F(\theta) d\theta,
\end{equation*}
\noindent where the densities are 
\begin{align*}
 f_G(x|\theta, \kappa) = \dfrac{\kappa}{\,\theta}\,x^{\, \kappa -1} \exp (-\dfrac{x^\kappa}{\theta})  \quad \quad f_F(\theta) = \dfrac{\beta^{\,\zeta}}{\Gamma(\zeta)\, \theta^{\,\zeta+1}} \exp(- \dfrac{\beta}{\theta}).
\end{align*}

The mean of distribution $H$ is given by
\begin{align*}
\mathbb{E}_{H}[X] =& \mathbb{E}_{F}[\mathbb{E}_{G}[X|\theta]]\\
=& \mathbb{E}_{F}[\theta^{1/k}\,\Gamma(1+\tfrac{1}{k})]\\
=& \int_{\theta} \theta^{1/k} \,\Gamma(1+\tfrac{1}{k}) \dfrac{\beta^{\, \zeta}}{\Gamma(\zeta) \theta^{\, \zeta + 1}} \exp(- \tfrac{\beta}{\theta}) d\theta\\
=& \dfrac{\Gamma(\zeta - \tfrac{1}{k}) \Gamma(1 + \tfrac{1}{k}) \beta^{1/k}} {\Gamma(\zeta) } \int_{\theta} \dfrac {\beta^{\, \zeta - 1/k}}{\Gamma(\zeta - \tfrac{1}{k}) \theta^{\, \zeta +1 -1/k}} \exp(-\tfrac{\beta}{\theta}) d\theta\\
=& \dfrac{\Gamma(\zeta - \tfrac{1}{k}) \Gamma(1 + \tfrac{1}{k}) \beta^{1/k}} {\Gamma(\zeta)}.
\end{align*}

This shows that the mean of the compound distribution of Weibull and Inverse-Gamma can be calculated in closed-form. Similarly, the variance of this compound distribution can also be obtained in closed-form as follows
\begin{align*}
    \textmd{Var}_{H}[X] =&  \mathbb{E}_F[\textmd{Var}_{G}[X|\theta]] + \textmd{Var}_{F}[\mathbb{E}_{G}[X|\theta]] \\
    =&  \dfrac{\Gamma(\zeta - \tfrac{2}{k}) [\Gamma(1 + \tfrac{2}{k}) -  (\Gamma(1 + \tfrac{1}{k}))^2] \beta^{2/k}} {\Gamma(\zeta)} + \dfrac{\Gamma(\zeta - \tfrac{2}{k}) (\Gamma(1 + \tfrac{1}{k}))^2 \beta^{2/k}} {\Gamma(\zeta)} - \mathbb{E}_{H}^2[X].
\end{align*}

%% file: Appendices/AppendixB.tex
\subsection{RDCEG Model Selection for the Falls Intervention}

\begin{figure}[!ht]
\centering
\includegraphics[trim = 0cm 1cm 0cm 0cm, scale = .33 ]{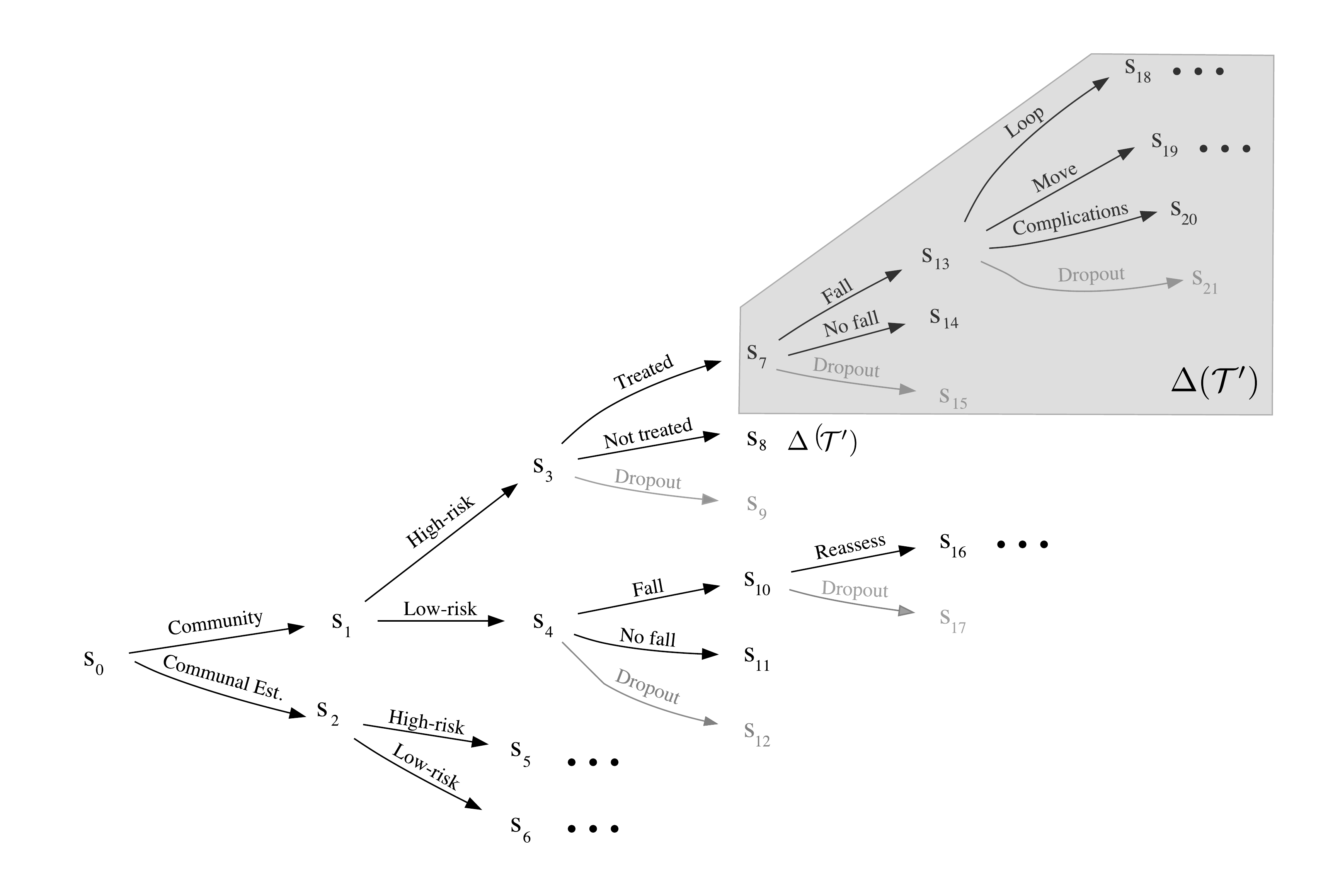}
\caption{The event tree for the falls intervention. The modified tree is given by the darkened vertices and edges. The subtrees emanating from situations $s_3$ and $s_5$ are identical with the exception of edge $e_{13,19}$. Similarly, the subtrees rooted at $s_4$ and $s_6$ are identical. The object representation of a subtree $\Delta(\calT')$ is used to denote repetition in structure.}
    \label{fig:event_tree}
\end{figure}

Figure \ref{fig:event_tree} shows the event tree for the falls intervention described in Section \ref{sec:simulation} of the article. It embeds the modeler's belief that low-risk individuals do not have complications arising directly from a fall and are reassessed for an updated fall-risk status after falling. The dropout vertices are $\{s_{9}, s_{12}, s_{15}, s_{17}, s_{21}, \ldots\}$. Vertex $s_{20}$ represents a critically terminating event. The darkened vertices and edges in Figure \ref{fig:event_tree} represent the modified tree $\calM$.  

We obtain the graph of the RDCEG $\calR$ presented in Figure \ref{fig:rdceg_falls} of the article by coloring the subtree of the hued tree $\calH$ corresponding to the first passage-slice $\calP_1$. We assume that the modeler's believe the falls to be independent up to a threshold number of falls. Under the assumptions of structural and parametric homogeneity, the data corresponding to passage-slices $\calP_k$, $k \geq 2$ can be used along with the data for $\calP_1$ to estimate the parameters of the RDCEG $\calR$.

After setting the priors, we then use the greedy Agglomerative Hierarchical Clustering algorithm in conjunction with collections called \textit{hyperstages} \citep{collazo_2017} and \textit{hyperclusters}. A hyperstage contains sets of situations which could be potentially hypothesized to be in the same stage. Similarly, a hypercluster contains sets of edges which could potentially be in the same cluster. These collections enable us to guide the search through the vast candidate model space in a structured way. For instance, there is little interpretative value in combining a situation corresponding to the type of residence of an individual to a situation concerned with the outcomes of a fall into one stage. In the case of the falls intervention, the hyperstages were set such that situations corresponding to the same variable can be in the same stage and hyperclusters were set to allow edges with the same known shape parameter to be in the same cluster. Note that, in our examples, we set the beta hyperparameter of the Inverse-Gamma prior distribution on the Weibull scale parameter to be equal for all edges and all clusters. This is done to reduce the influence of this prior as it is set to be weakly informative.

\begin{figure}[!ht]
\begin{subfigure}{0.30\textwidth}
\centering
  \includegraphics[trim = 0cm 0cm 0cm 0cm, scale = .26 ]{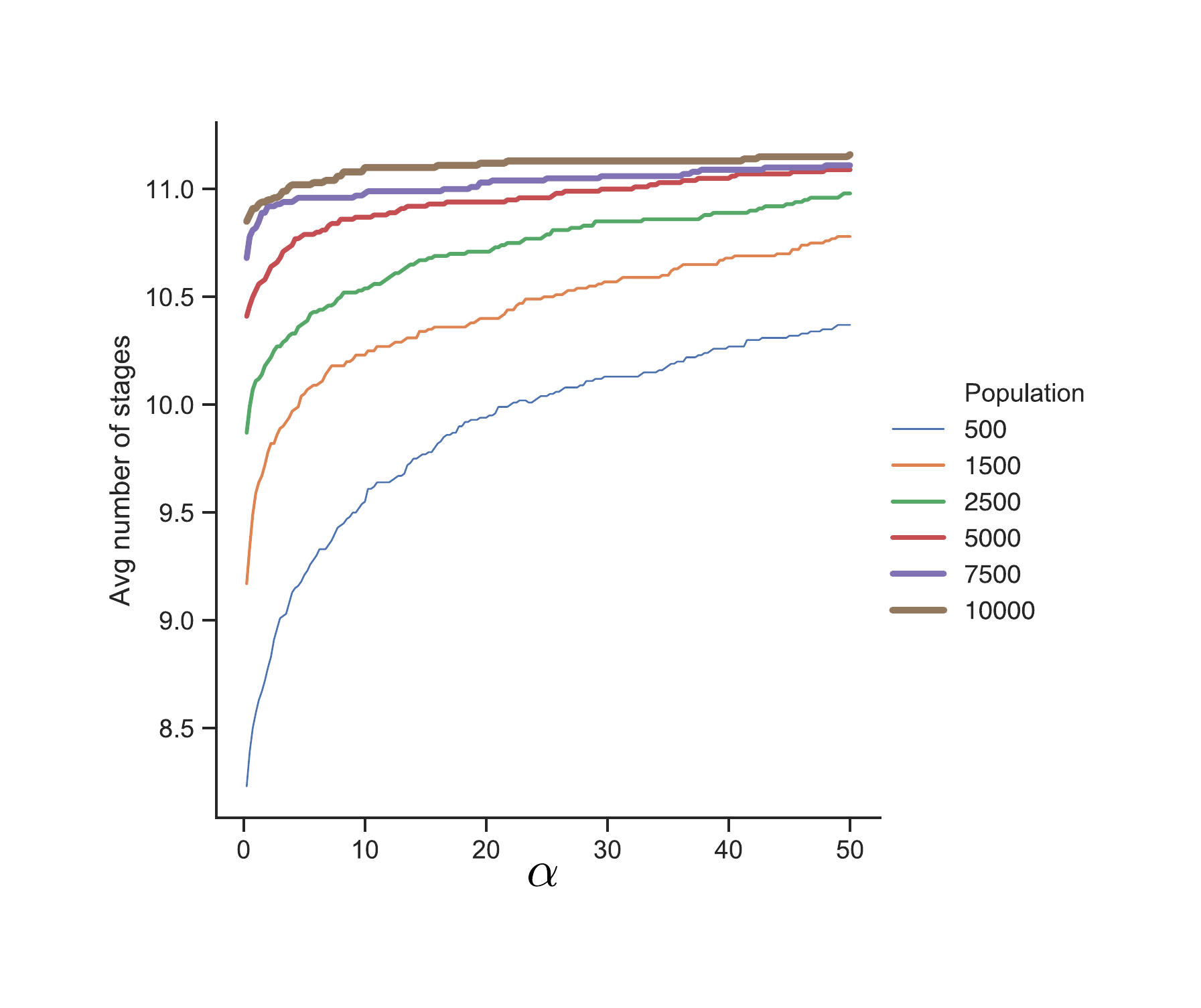}
  \caption{Average number of stages}
    \label{fig:avg_stages}
\end{subfigure}
\begin{subfigure}{0.68\textwidth}
\centering
  \includegraphics[trim = 0cm 0cm 0cm 0cm, scale = .31 ]{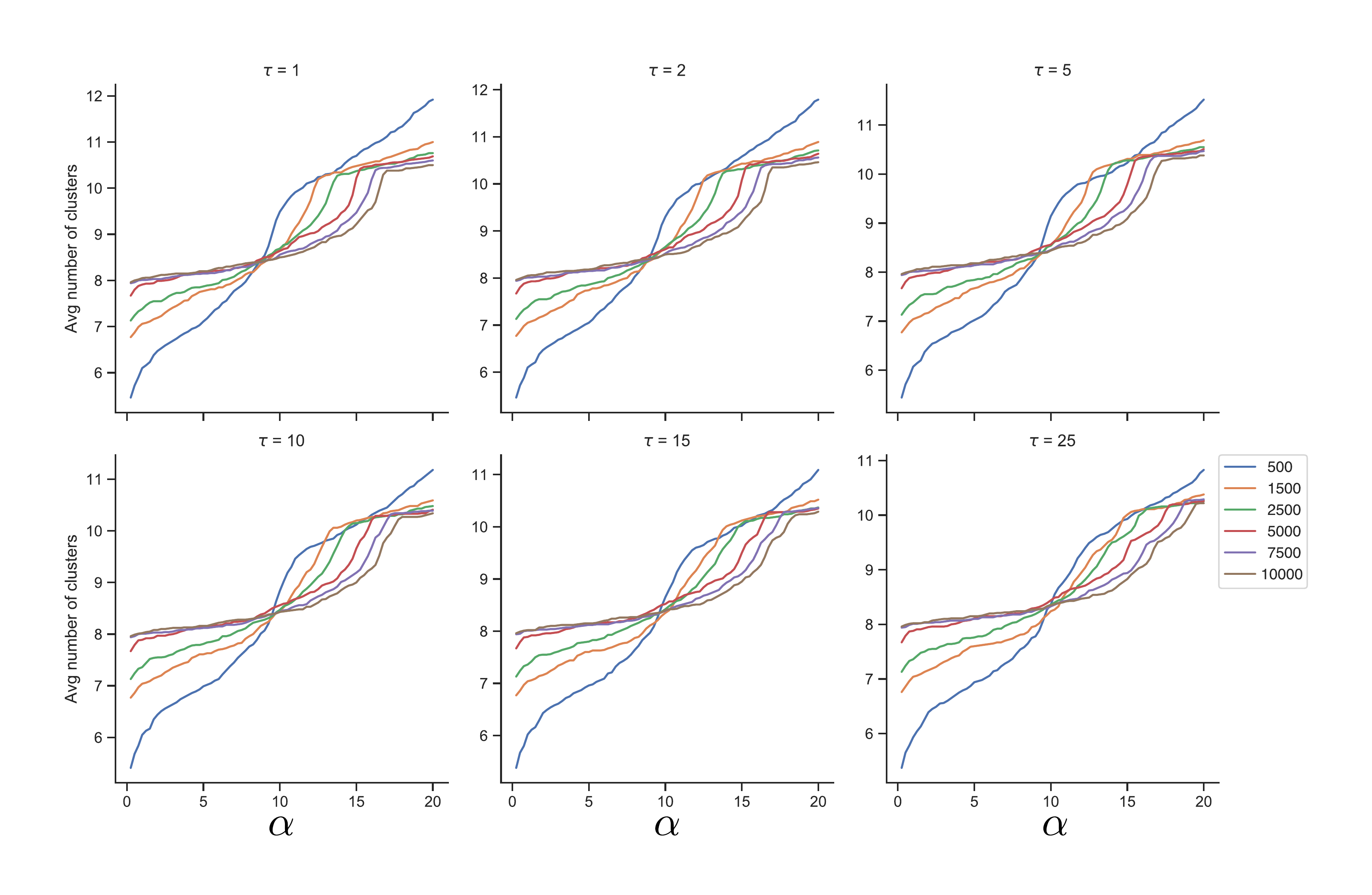}
  \caption{Average number of clusters}
    \label{fig:avg_clusters}
\end{subfigure}
\caption{Averages taken over 100 simulations for each sample size. $\tau$ denotes the phantom holding time and $\alpha$ is the total number of phantom units.}
\end{figure}

Figures \ref{fig:avg_stages} and \ref{fig:avg_clusters} shows the average number of stages and clusters across the simulations for the different prior specifications and different population sizes. We see that the average number of stages stabilize quickly as the population size increases. The average number of clusters are more volatile as they depend on two parameters. From Figure \ref{fig:avg_clusters}, we can see that the average cluster size is more stable for smaller values of $\alpha$ which is the total number of phantom units affecting the shape hyperparameter of the Inverse-Gamma prior distributions. 

\begin{figure}[!ht]
\begin{subfigure}{0.49\textwidth}
\centering
  \includegraphics[trim = 5cm 0cm 5cm 0cm, scale = .22 ]{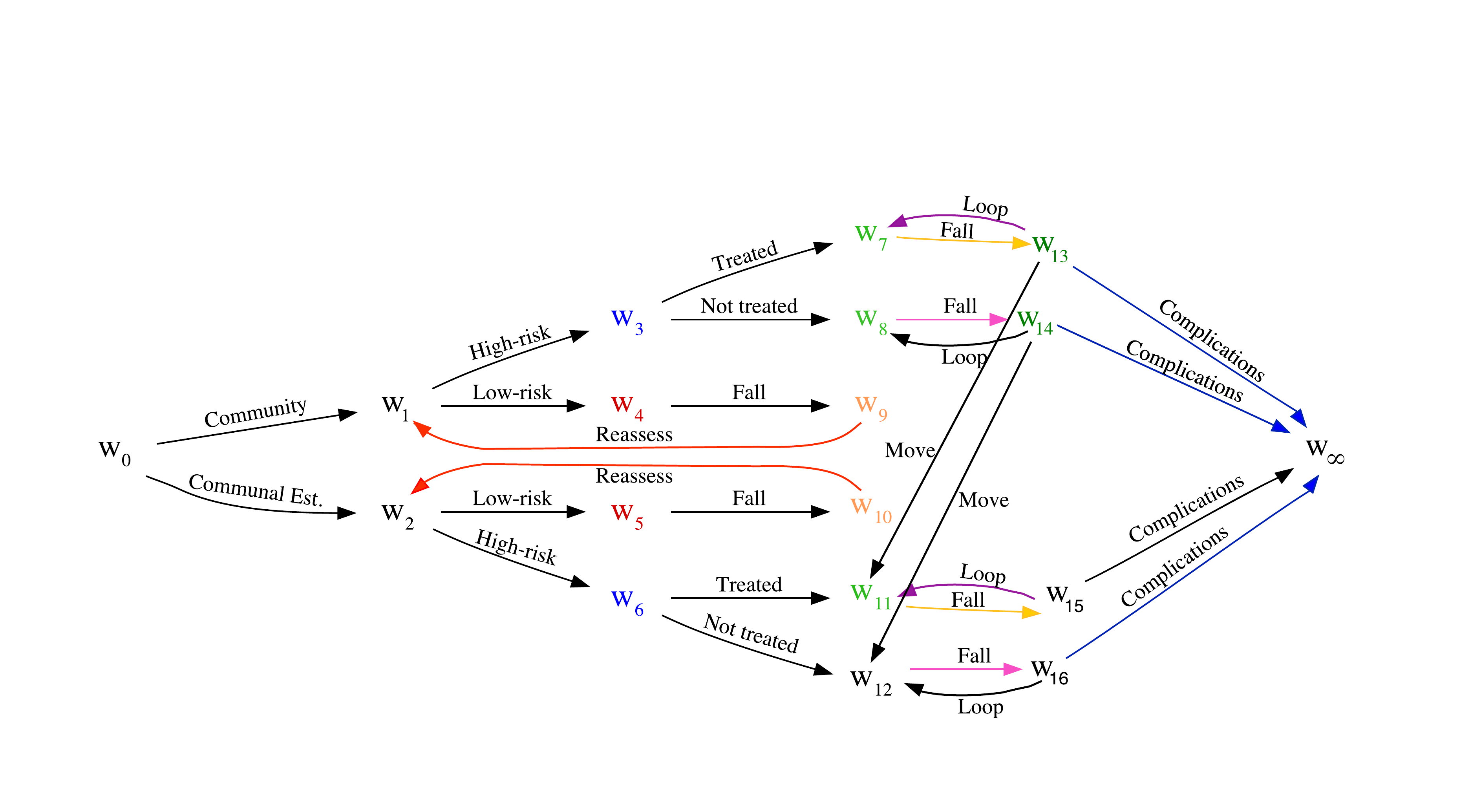}
  \caption{}
    \label{fig:alt_a}
\end{subfigure}
\begin{subfigure}{0.49\textwidth}
\centering
  \includegraphics[trim = 5cm 0cm 5cm 0cm, scale = .22 ]{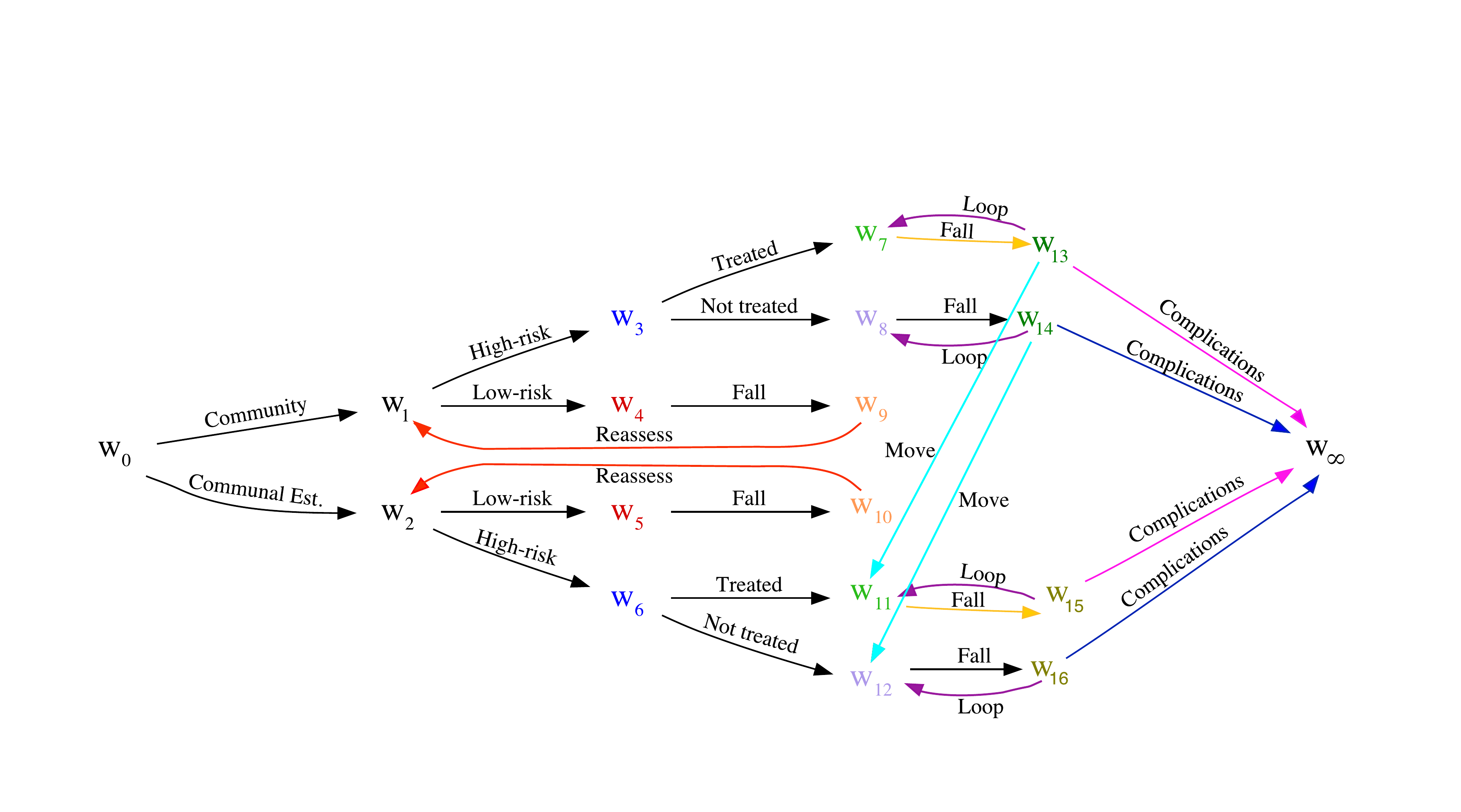}
  \caption{}
    \label{fig:alt_b}
\end{subfigure}
\newline
\begin{subfigure}{\textwidth}
\centering
  \includegraphics[trim = 5cm 0cm 5cm 0cm, scale = .23 ]{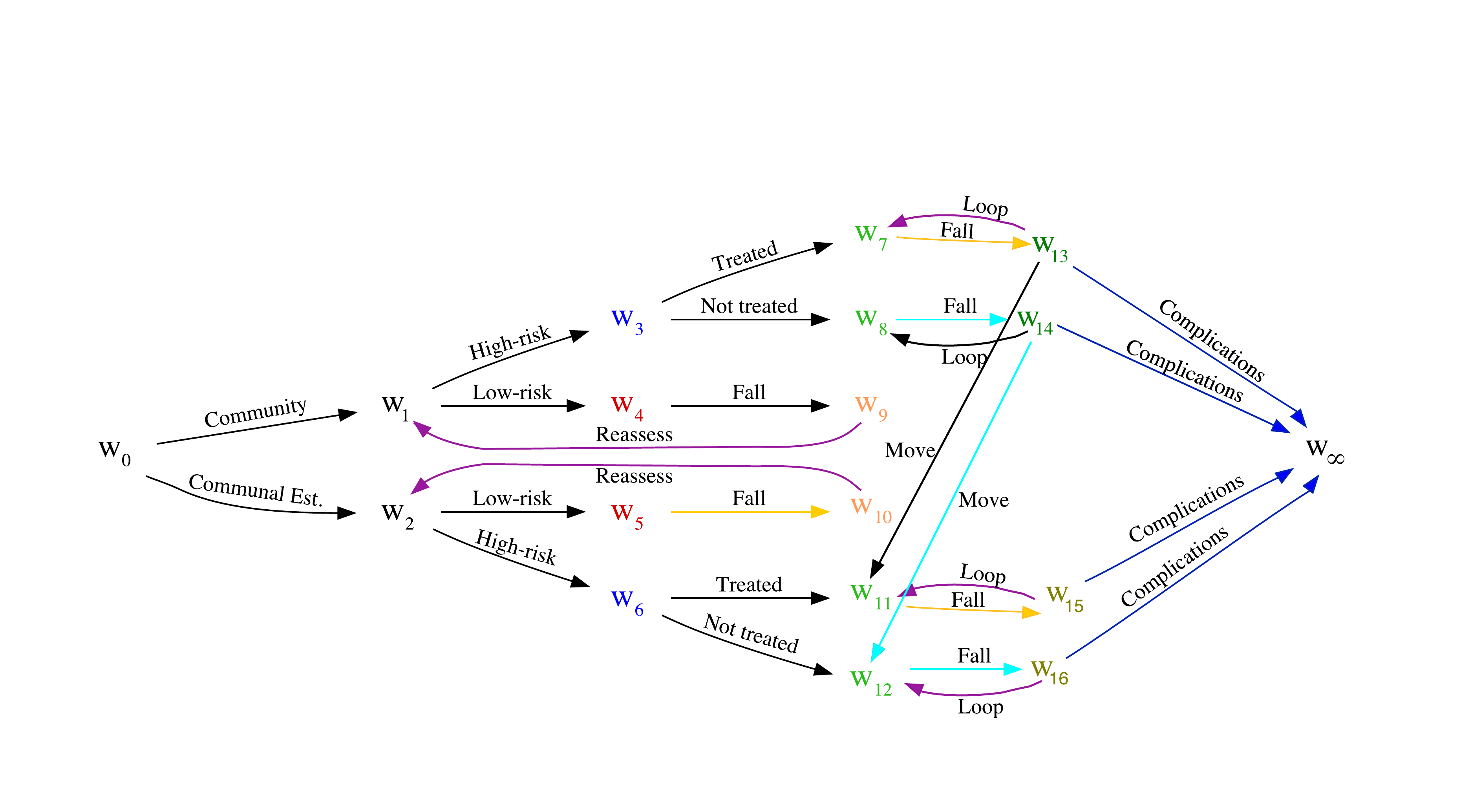}
  \caption{}
    \label{fig:alt_c}
\end{subfigure}
\caption{Alternative MAP RDCEG models found by the Agglomerative Hierarchical Clustering algorithm.}
\label{fig:alternative_rdcegs}
\end{figure}

Figure \ref{fig:alternative_rdcegs} gives some of the alternative MAP RDCEGs - found by our model selection algorithm for the different population sizes and prior specifications - which appear to be quite different from our generating RDCEG . While these RDCEGs typically appeared only for the low population size of 500 and that too, for extreme prior specifications, we will briefly analyze the conclusions we can draw from them. Recall that these conclusions are valid to individuals conditional on them not dropping out of the population. The RDCEG in Figure \ref{fig:alt_a} shows that the probability of falling for community-dwelling high-risk individuals is the same as that for all the individuals who are treated. Figure \ref{fig:alt_b} indicates a difference in the duration from a fall to exiting the population due to serious complications for high-risk individuals who receive treatment and those who do not. It also shows that the probability of falling is the same for all high-risk individuals who are not treated. The models in Figures \ref{fig:alt_a} and \ref{fig:alt_b} are, however, not very different from our generating RDCEG model. The model given by the graph in Figure \ref{fig:alt_c} shows that the probability of falling is the same for all high-risk individuals irrespective of whether they have received any treatment. It also shows that the duration from assessment to a fall is the same for all high-risk treated individuals and for low-risk individuals living in communal establishments. This model is significantly different from our generating model. However, we found that it was selected by the model selection algorithm only for the population size of 500 and for a limited range of prior specifications.

%% file: Appendices/AppendixC.tex
\subsection{Diagnostics for the Simplified RDCEG of the Epilepsy Trial}

The conclusions we can draw from routine model selection methods are less reliable when the number of individuals presented in a certain subgroup is small. In the Epilepsy case study presented in Section \ref{sec:case} of the article, there are fewer than 30 individuals with unknown EEG results for each age group. Most of our data (around 94\%) concerns those with known EEG results. So we extended and applied some CEG diagnostic monitors to check our model fit. 

\begin{figure}[h!]
\centering
  \includegraphics[trim = 0cm 0cm 0cm 5cm, width = 0.45\textwidth, height = 0.8\textheight]{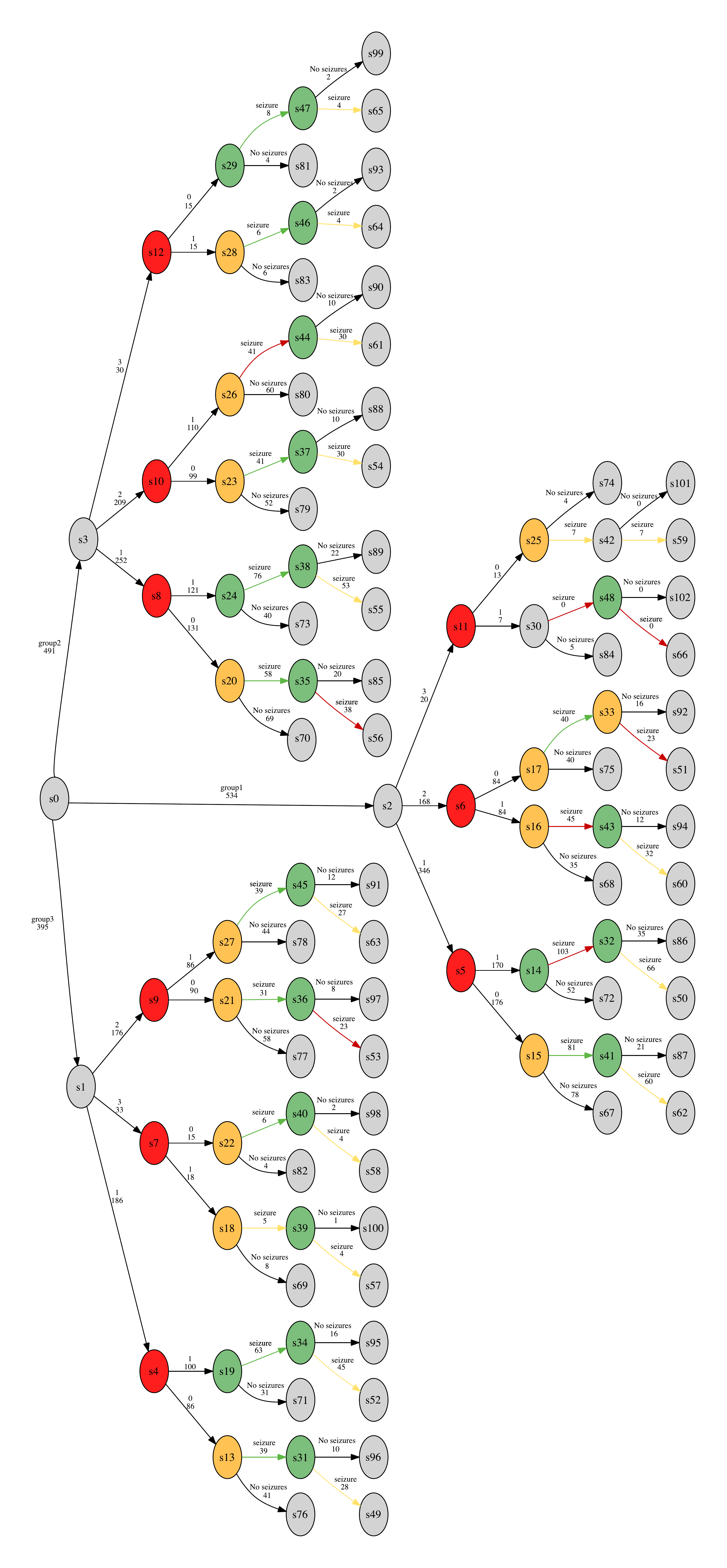}
\caption{The hued tree for the epilepsy case study. The edges emanating from $s_i$, $i = 1,2,3$ are labeled as `1' for an abnormal EEG, `2' for a normal EEG and `3' for an unknown EEG result. The edges emanating from $s_i$, $i = 4,\ldots, 12$ are labeled as `0' for immediate treatment and `1' for deferred treatment.}
\label{fig:rdceg_colored}
\end{figure}

\begin{figure}[!ht]
\centering
\begin{subfigure}[b]{0.31 \textwidth}
\centering
\includegraphics[trim = 0cm 0cm 0cm 0cm, scale = 0.33]{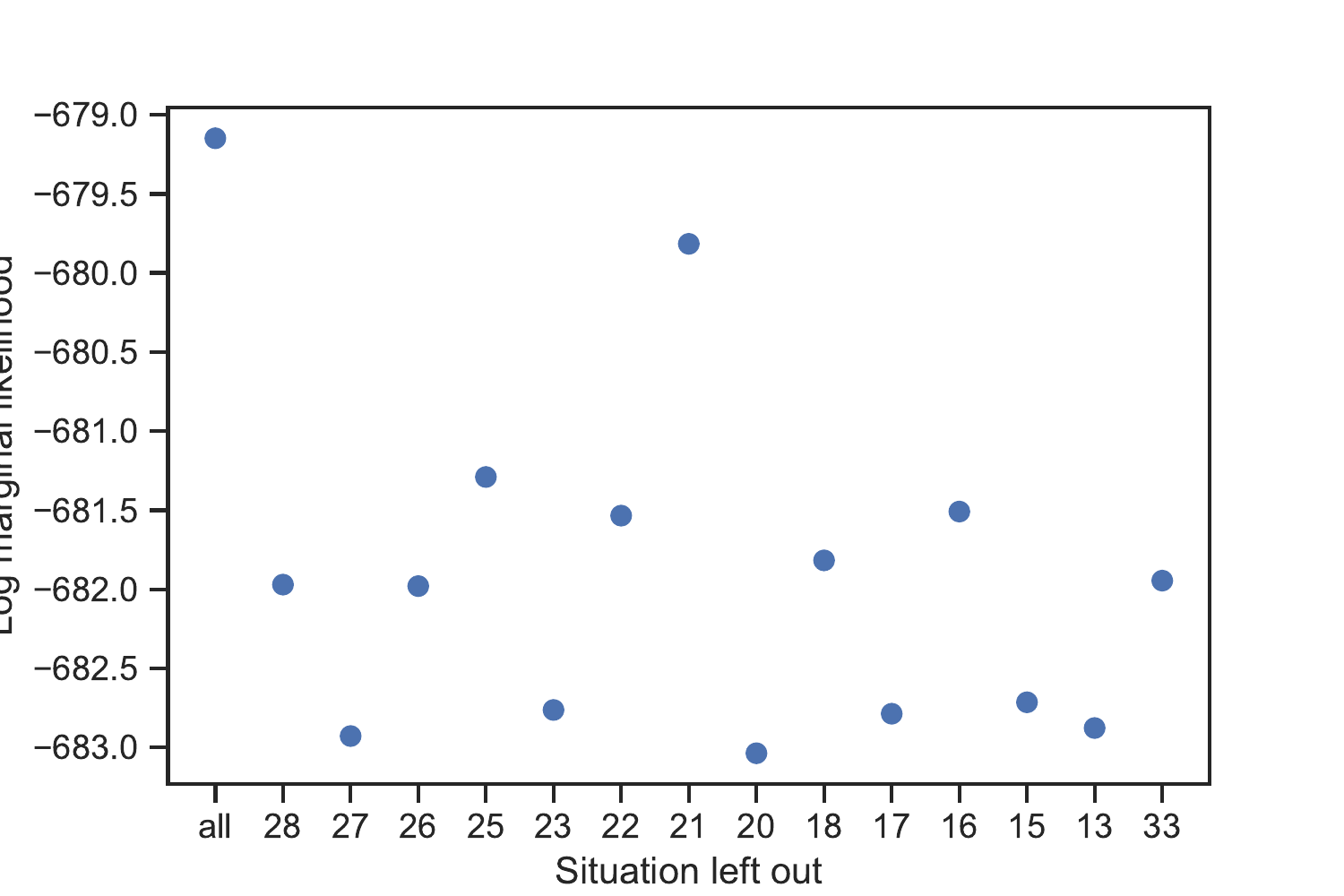}
\caption{}
\end{subfigure}
\begin{subfigure}[b]{0.31\textwidth}
\centering
\includegraphics[trim = 0cm 0cm 0cm 0cm, scale = 0.33]{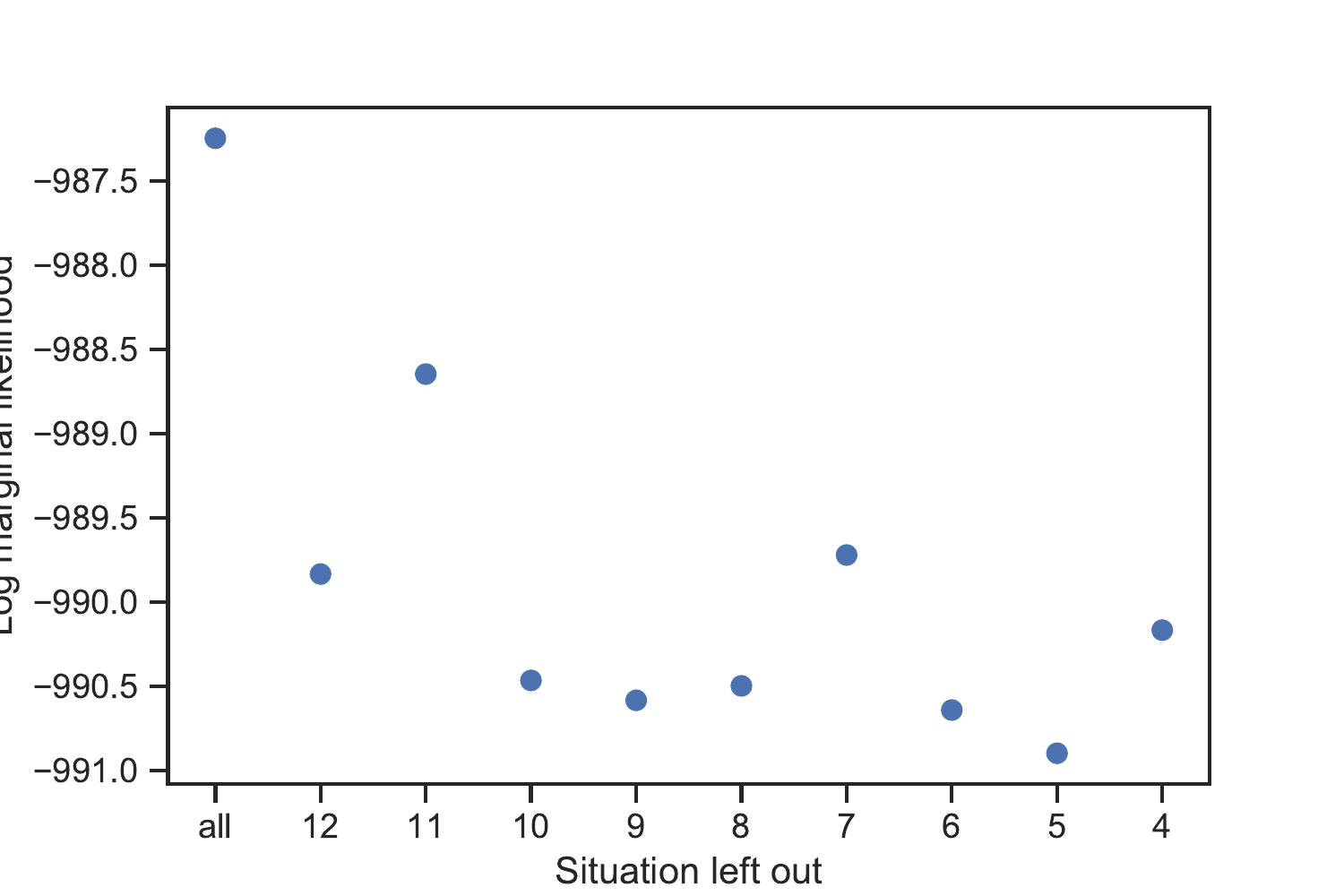}
\caption{}
\end{subfigure}
\begin{subfigure}[b]{0.31\textwidth}
\centering
\includegraphics[trim = 0cm 0cm 0cm 0cm, scale = 0.33]{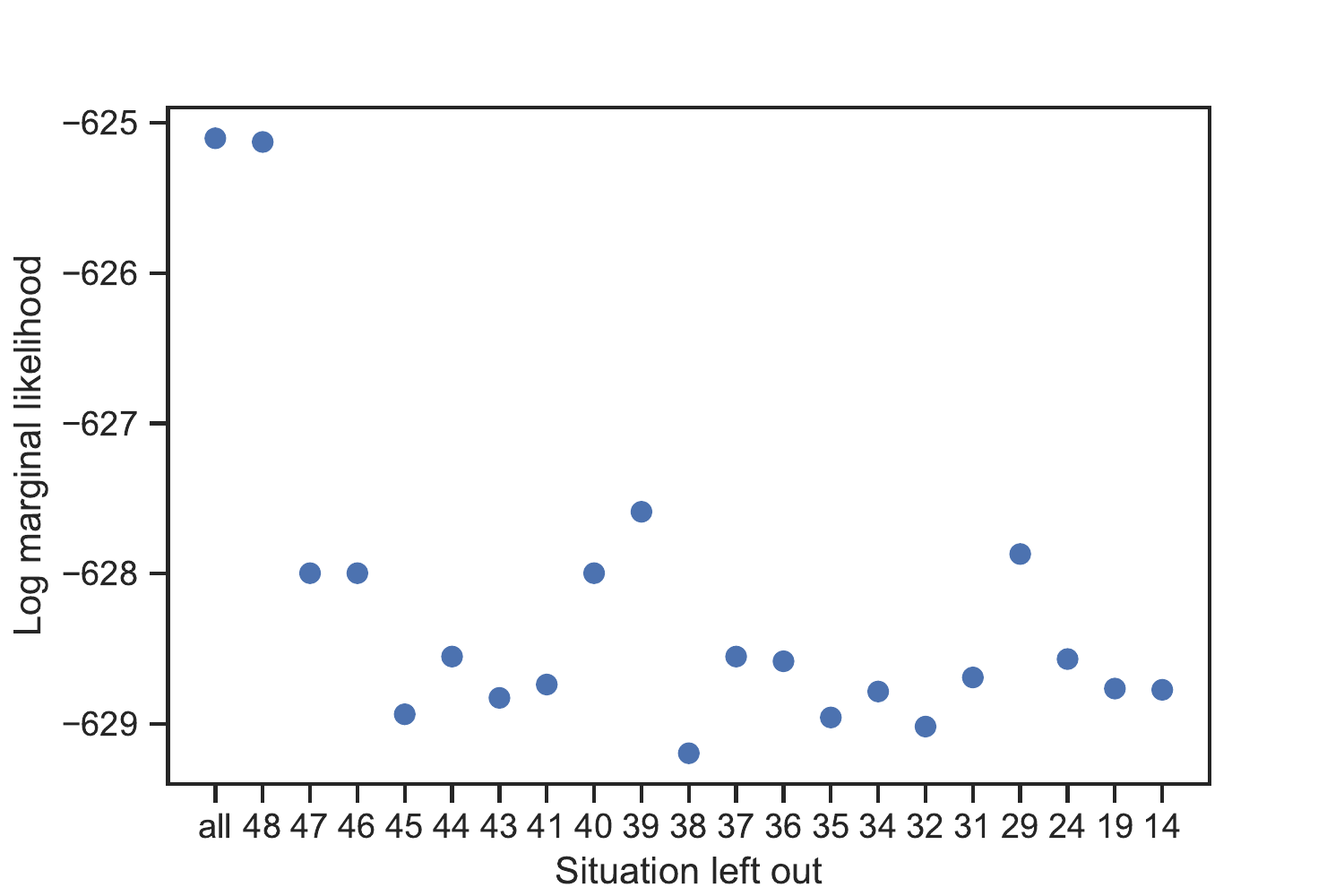}
\caption{}
\end{subfigure}
\newline
\begin{subfigure}[b]{0.31\textwidth}
\centering
\includegraphics[trim = 0cm 0cm 0cm 0cm, scale = 0.33]{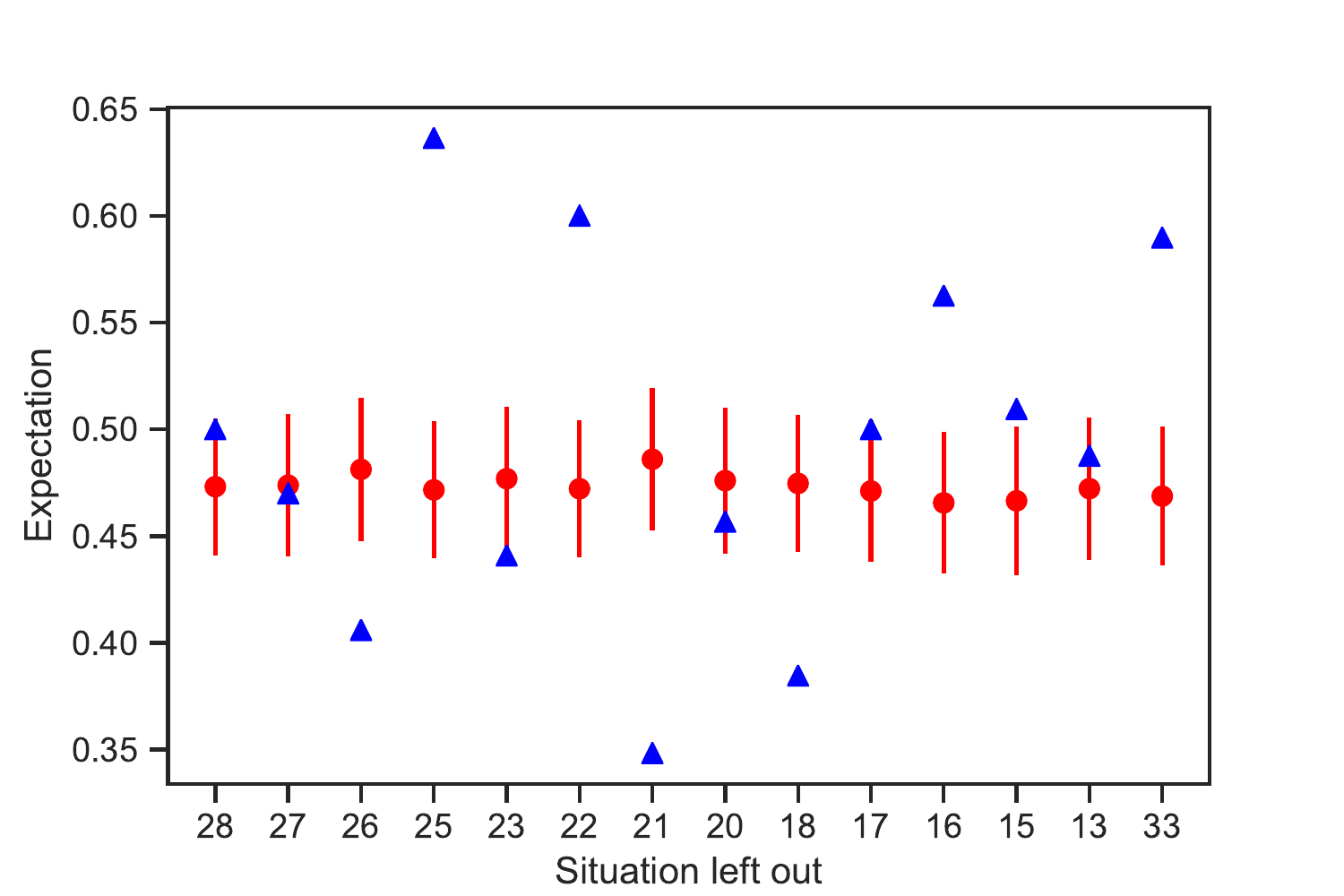}
\caption{}
\end{subfigure}
\begin{subfigure}[b]{0.31\textwidth}
\centering
\includegraphics[trim = 0cm 0cm 0cm 0cm, scale = 0.33]{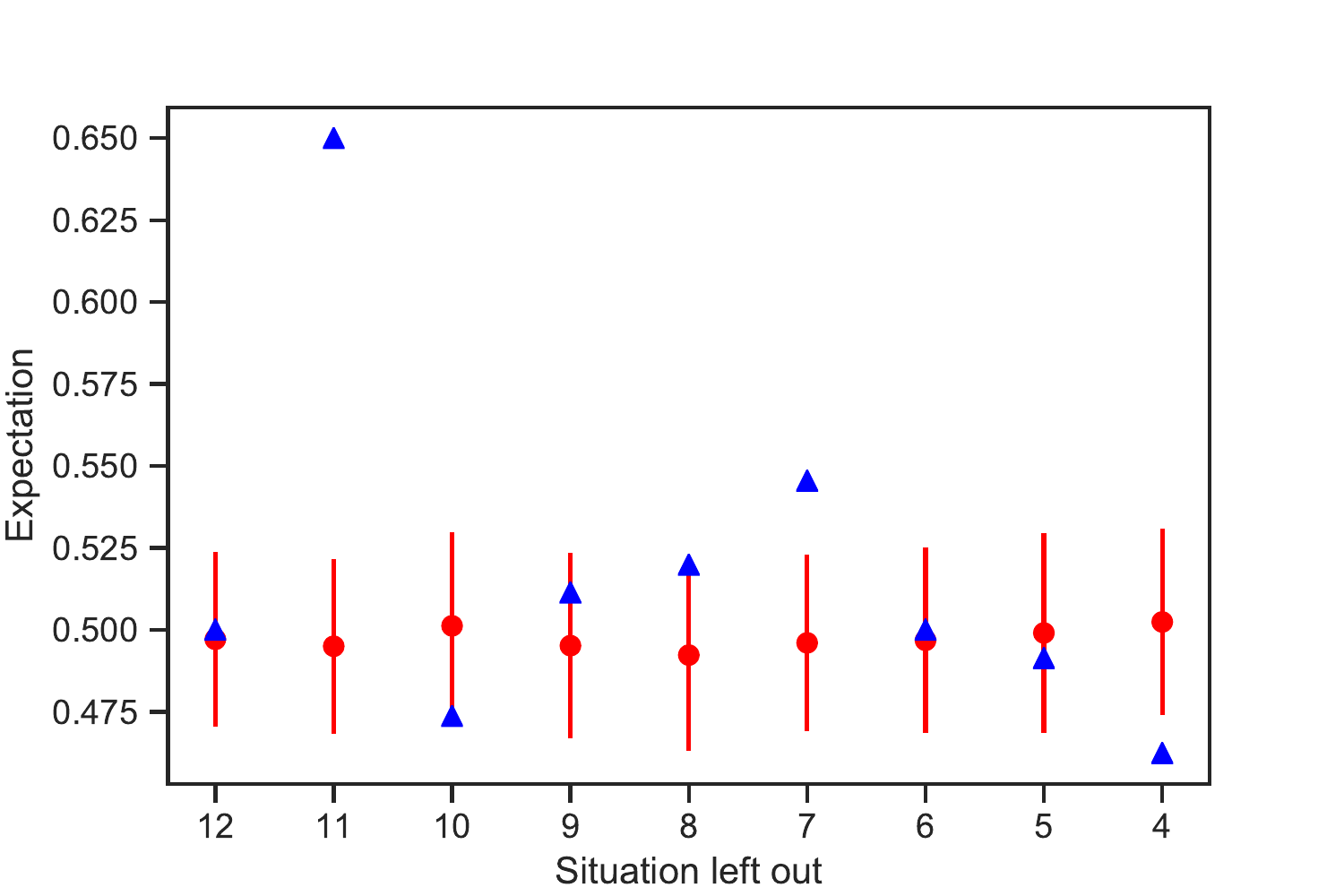}
\caption{}
\end{subfigure}
\begin{subfigure}[b]{0.31\textwidth}
\centering
\includegraphics[trim = 0cm 0cm 0cm 0cm, scale = 0.33]{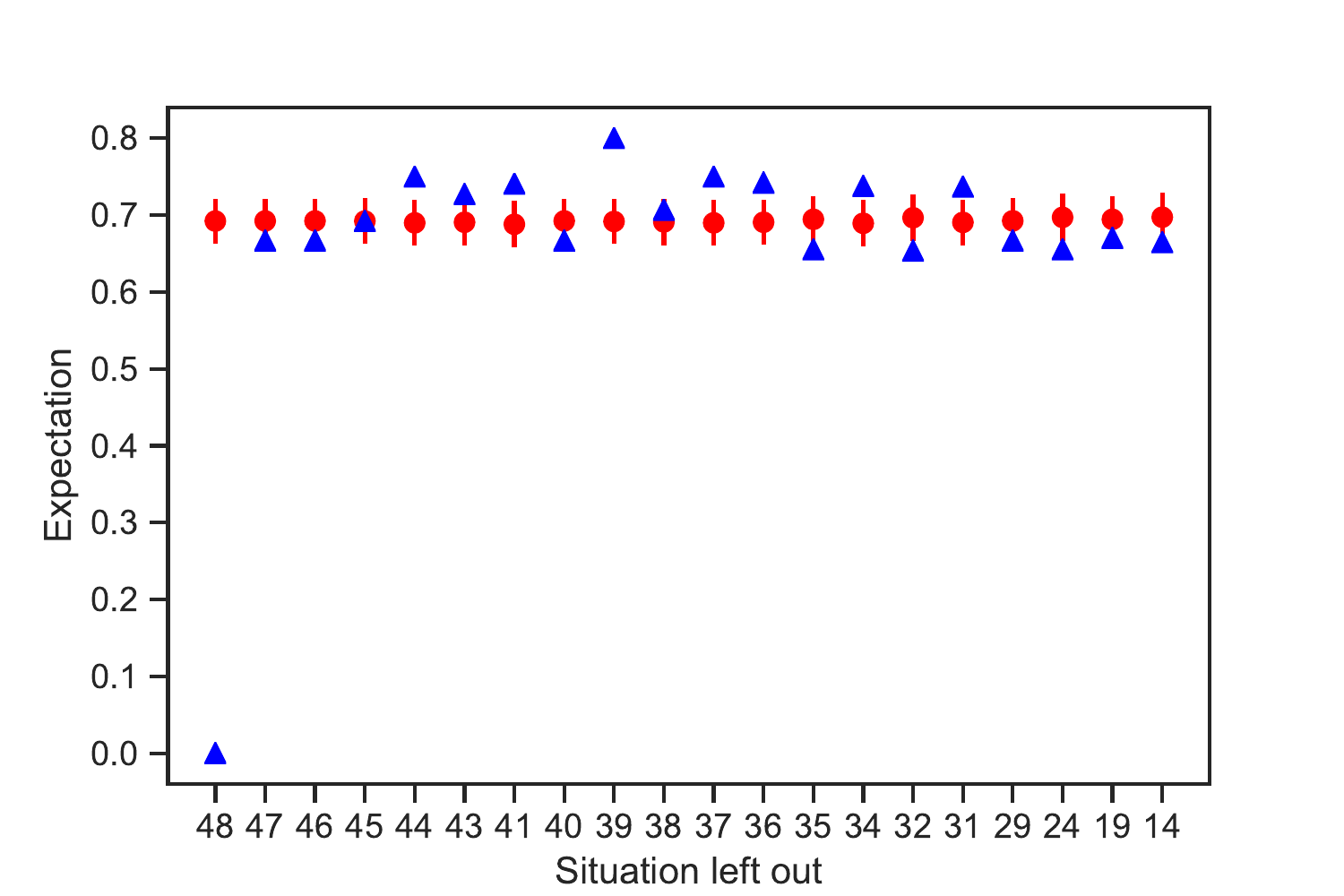}
\caption{}
\end{subfigure}
\caption{The x-axis indicates the index of the situation left out. This indexing corresponds to the indexing of the situations in Figure \ref{fig:rdceg_colored}. `All' indicates that no situation has been left out. Figures (a-c) show the total log marginal likelihood score of the situation taken out of the current stage and of the stage without that situation. In Figures (d-f), the red dots show the posterior expectation and the red lines indicate two standard deviations from this expectation of the stage after leaving a situation out. The blue triangles show the observed mean of the situation that has been left out.}
    \label{fig:loo_stages}
\end{figure}

\begin{figure}[!ht]
\centering
\begin{subfigure}[b]{0.31 \textwidth}
\centering
\includegraphics[trim = 0cm 0cm 0cm 0cm, scale = 0.33]{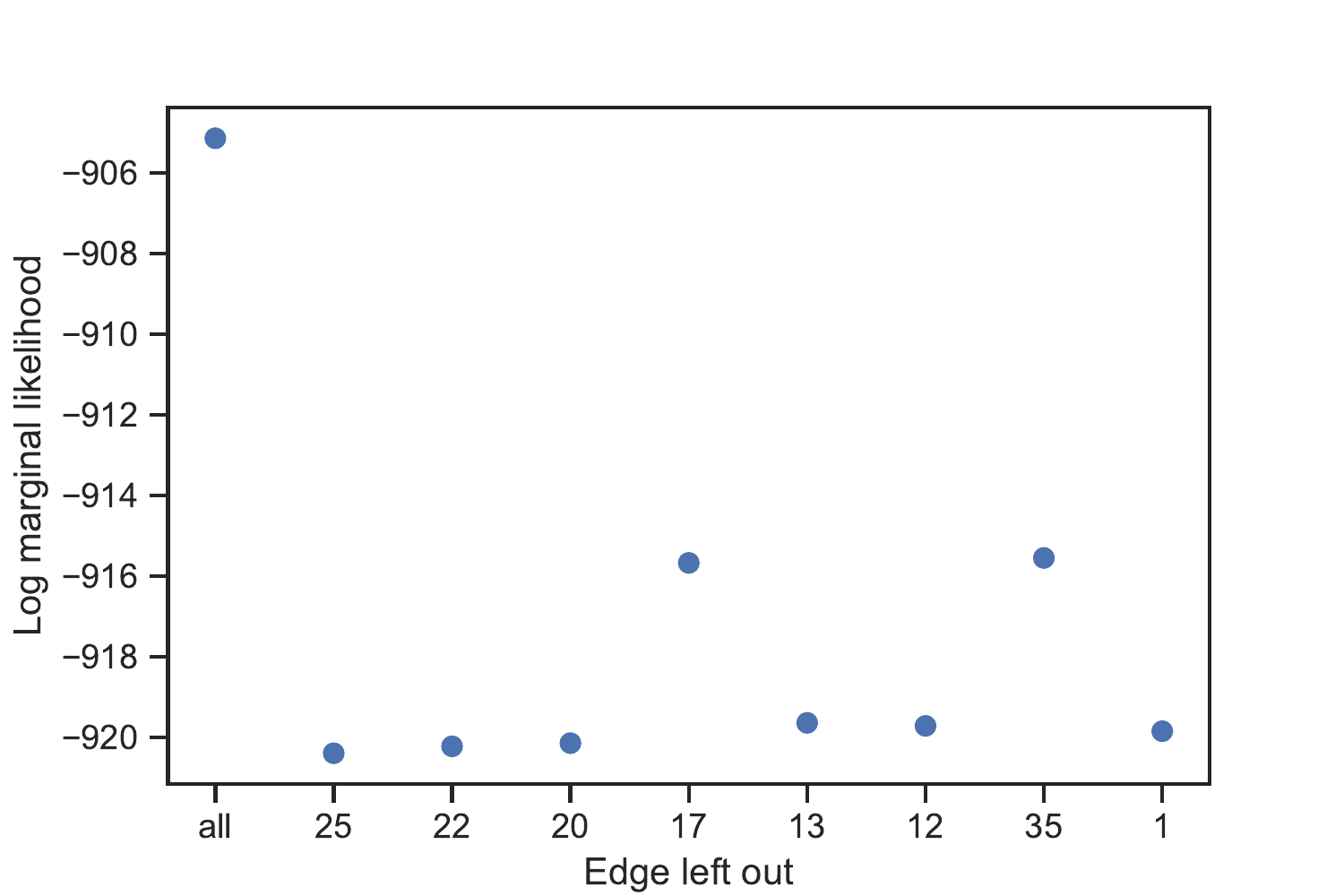}
\caption{}
\end{subfigure}
\begin{subfigure}[b]{0.31\textwidth}
\centering
\includegraphics[trim = 0cm 0cm 0cm 0cm, scale = 0.33]{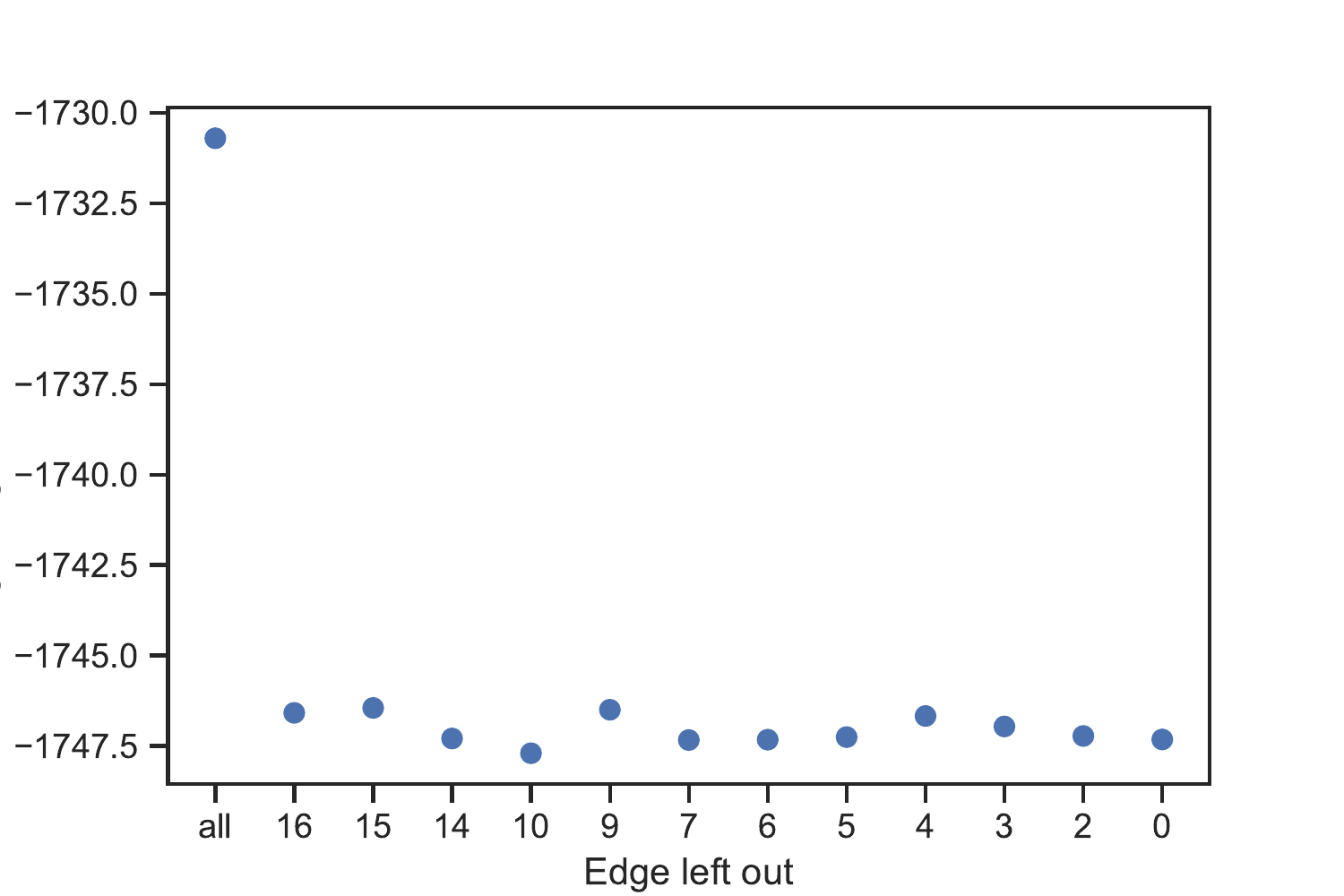}
\caption{}
\end{subfigure}
\begin{subfigure}[b]{0.31\textwidth}
\centering
\includegraphics[trim = 0cm 0cm 0cm 0cm, scale = 0.33]{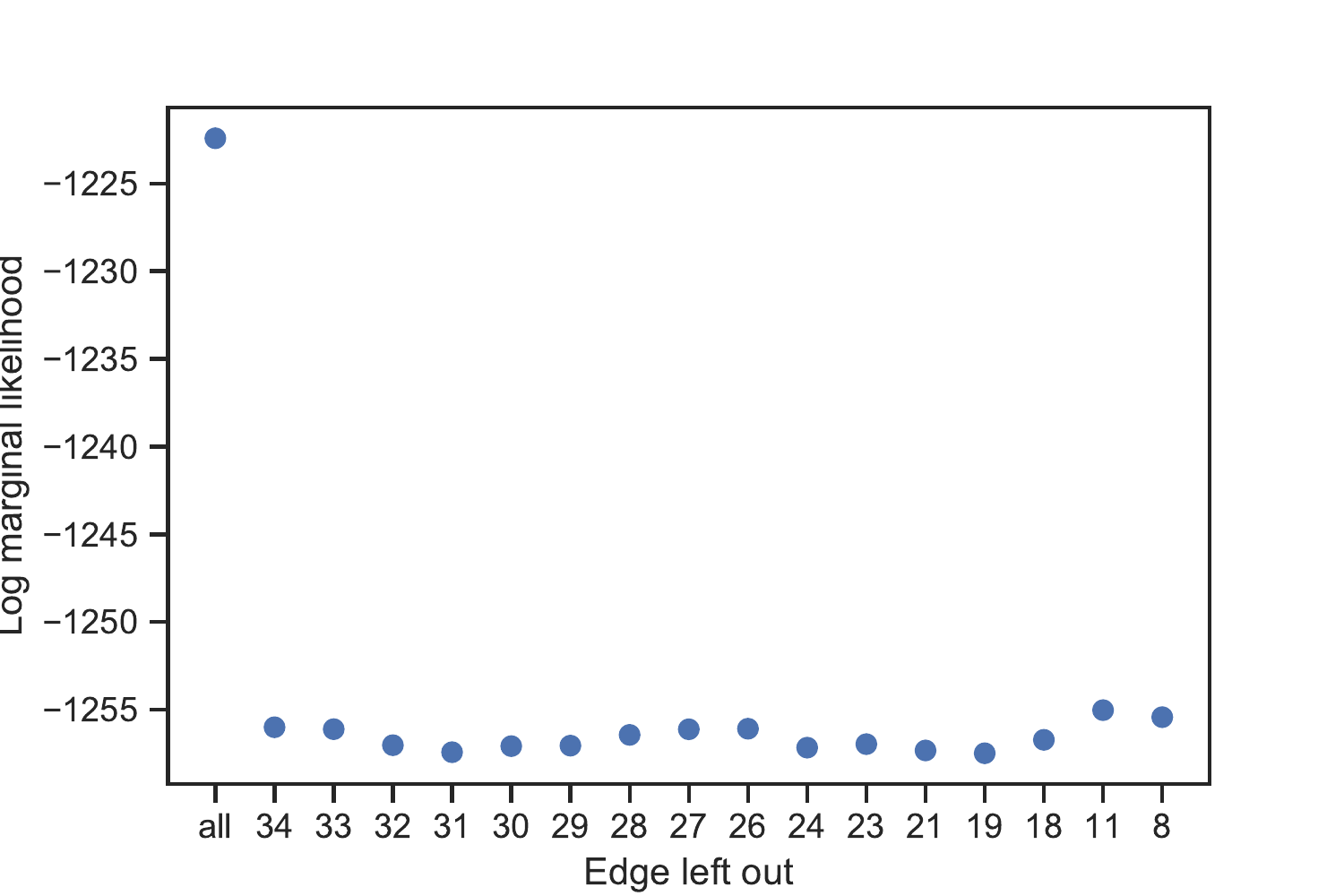}
\caption{}
\end{subfigure}
\newline
\begin{subfigure}[b]{0.31\textwidth}
\centering
\includegraphics[trim = 0cm 0cm 0cm 0cm, scale = 0.33]{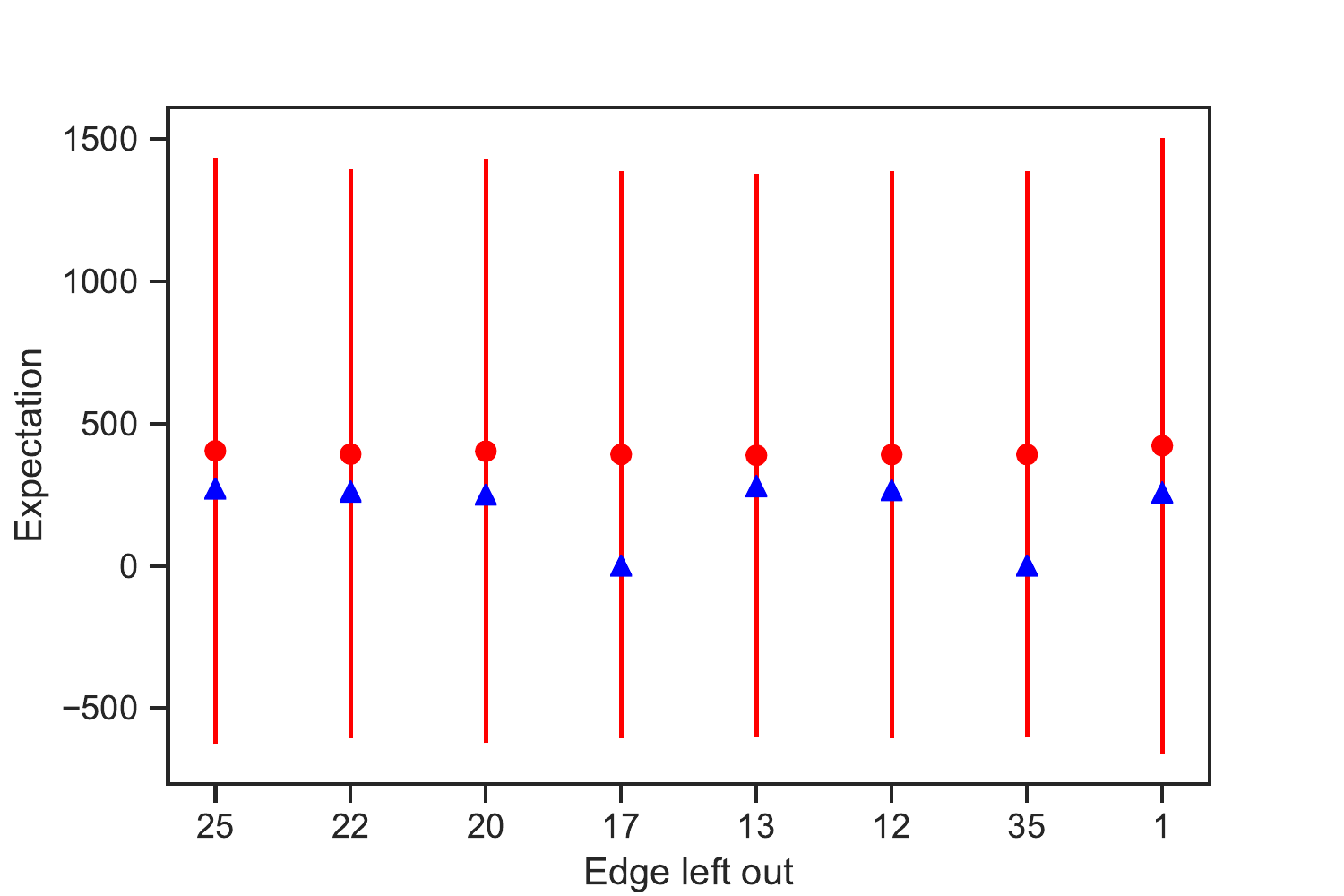}
\caption{}
\end{subfigure}
\begin{subfigure}[b]{0.31\textwidth}
\centering
\includegraphics[trim = 0cm 0cm 0cm 0cm, scale = 0.33]{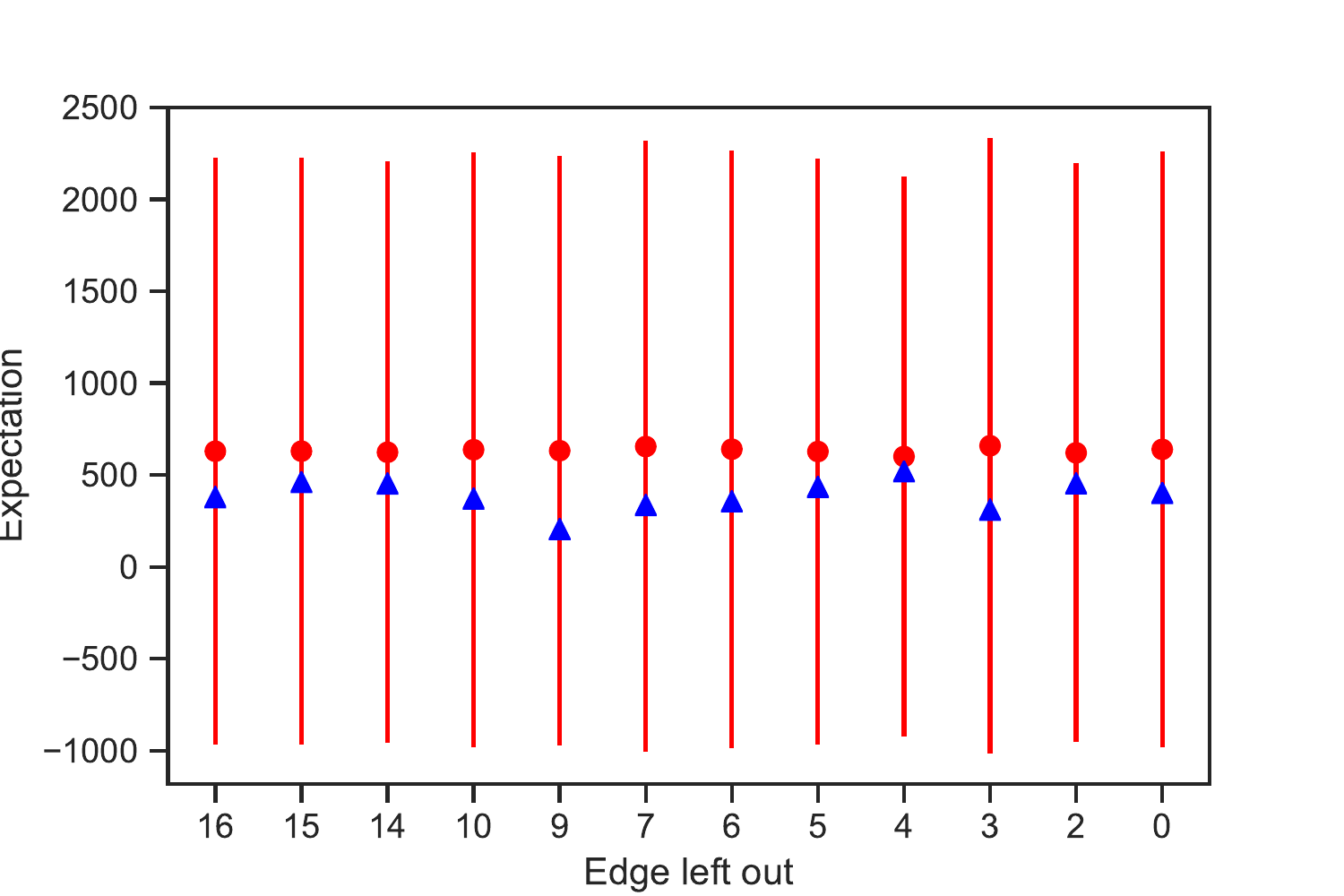}
\caption{}
\end{subfigure}
\begin{subfigure}[b]{0.31\textwidth}
\centering
\includegraphics[trim = 0cm 0cm 0cm 0cm, scale = 0.33]{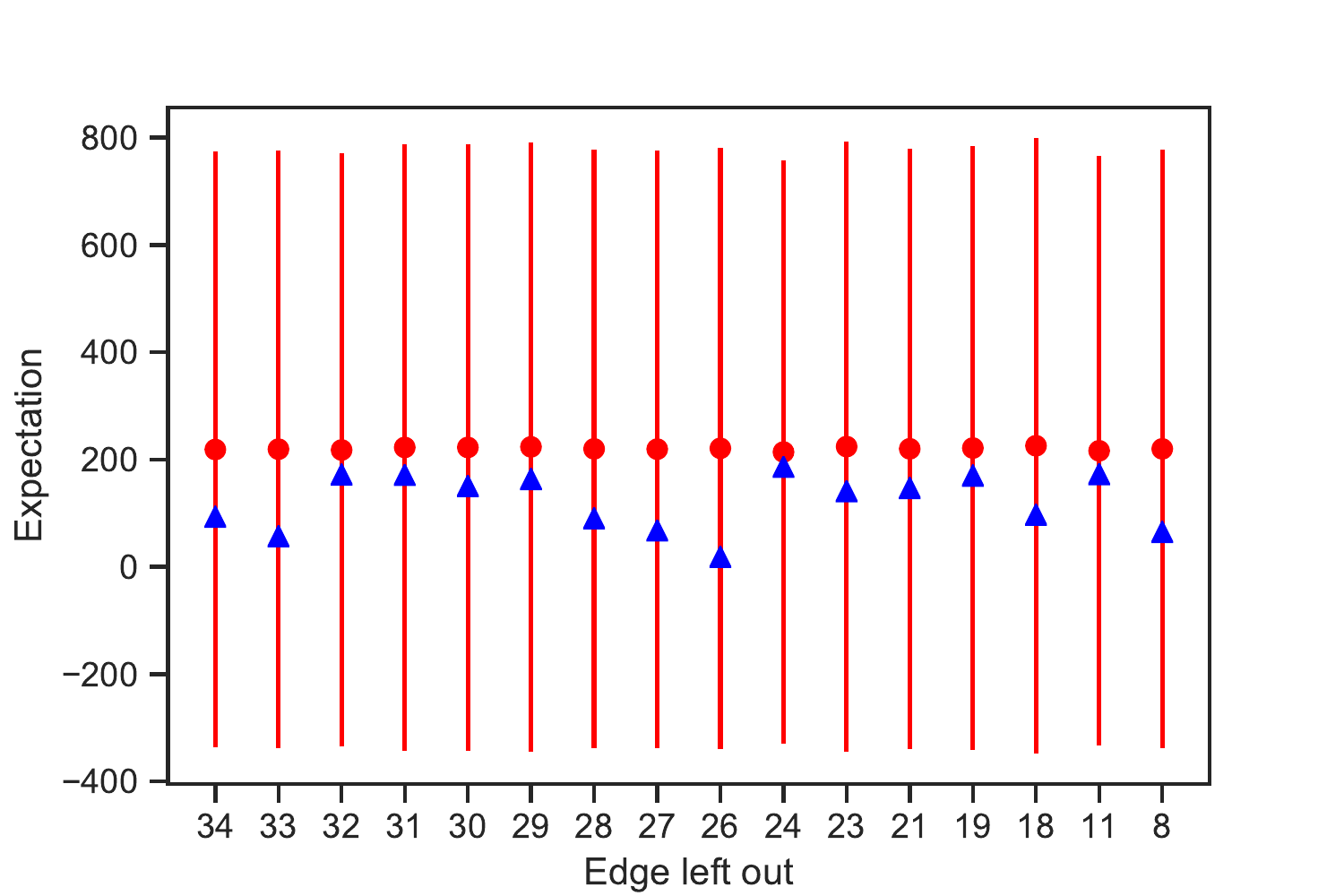}
\caption{}
\end{subfigure}
\caption{The x-axis indicates the number of the edge left out (refer to Table \ref{table:edge_key}). `All' indicates that no edge has been left out. Figures (a-c) show the total log marginal likelihood score of the edge taken out of the current cluster and of the cluster without that edge. In Figures (d-f), the red dots show the posterior expectation (in number of days) and the red lines indicate one standard deviation from this expectation of the cluster after leaving an edge out. The blue triangles show the observed mean of the edge that has been left out.}
    \label{fig:loo_clusters}
\end{figure}

We now apply the leave-one-out diagnostic monitor presented in \cite{wilkersonbayesian2019} to the RDCEG chosen by the model selection algorithm for the epilepsy case study in Section \ref{sec:case} of the article. As this monitor was initially developed for CEGs which did not have any differential holding times on the edges, we adapted it to apply to clusters in the RDCEG. Figure \ref{fig:rdceg_colored} shows the hued tree for the epilepsy case study. Figures \ref{fig:loo_stages} and \ref{fig:loo_clusters} visualize the results of the leave-one-out monitor for stages and clusters respectively. This monitor indicates whether a particular situation or edge is a bad fit to the stage or cluster it has been assigned by the model search algorithm. 

\begin{table}[h!]
\centering
\resizebox{0.32\linewidth}{!}{%
\begin{tabular}{|c|l|}
\hline
\textbf{Edge number} &  \textbf{Path to edge} \\ \hline
 0 &           (group3, 1, 0, seizure) \\ \hline
 1 &           (group1, 1, 1, seizure) \\ \hline
 2 &           (group1, 2, 0, seizure) \\ \hline
 3 &           (group3, 1, 1, seizure) \\ \hline
 4 &           (group2, 1, 0, seizure) \\ \hline
 5 &           (group3, 2, 0, seizure) \\ \hline
 6 &           (group2, 2, 0, seizure) \\ \hline
 7 &           (group2, 1, 1, seizure) \\ \hline
 8 &           (group3, 3, 1, seizure) \\ \hline
 9 &           (group3, 3, 0, seizure) \\ \hline
10 &           (group1, 1, 0, seizure) \\ \hline
11 &           (group1, 3, 0, seizure) \\ \hline
12 &           (group1, 2, 1, seizure) \\ \hline
13 &           (group2, 2, 1, seizure) \\ \hline
14 &           (group3, 2, 1, seizure) \\ \hline
15 &           (group2, 3, 1, seizure) \\ \hline
16 &           (group2, 3, 0, seizure) \\ \hline
17 &           (group1, 3, 1, seizure) \\ \hline
18 &  (group3, 1, 0, seizure, seizure) \\ \hline
19 &  (group1, 1, 1, seizure, seizure) \\ \hline
20 &  (group1, 2, 0, seizure, seizure) \\ \hline
21 &  (group3, 1, 1, seizure, seizure) \\ \hline
22 &  (group3, 2, 0, seizure, seizure) \\ \hline
23 &  (group2, 2, 0, seizure, seizure) \\ \hline
24 &  (group2, 1, 1, seizure, seizure) \\ \hline
25 &  (group2, 1, 0, seizure, seizure) \\ \hline
26 &  (group3, 3, 1, seizure, seizure) \\ \hline
27 &  (group3, 3, 0, seizure, seizure) \\ \hline
28 &  (group1, 3, 0, seizure, seizure) \\ \hline
29 &  (group1, 2, 1, seizure, seizure) \\ \hline
30 &  (group2, 2, 1, seizure, seizure) \\ \hline
31 &  (group1, 1, 0, seizure, seizure) \\ \hline
32 &  (group3, 2, 1, seizure, seizure) \\ \hline
33 &  (group2, 3, 1, seizure, seizure) \\ \hline
34 &  (group2, 3, 0, seizure, seizure) \\ \hline
35 &  (group1, 3, 1, seizure, seizure) \\ \hline
\end{tabular}
}
\caption{Each edge, indicated by the final edge on the path given in the second column, is identified by its corresponding number in the first column in Figure \ref{fig:loo_clusters}.}
\label{table:edge_key}
\end{table}

The log marginal likelihood scores in Figure \ref{fig:loo_stages}(a-c) show that the stages created by the model selection algorithm score highest when none of their constituent situations are left out. However, there are several situations whose observed means do not fall within two standard deviations of the posterior mean of their stages without them in it. This indicates that we should look into these situations closer. Most of these situations picked out by the leave-one-out monitor ($s_7$, $s_{11}$, $s_{18}$, $s_{22}$, $s_{25}$, $s_{39}$ and $s_{48}$) have very few observations along their emanating edges as can be seen in Figure \ref{fig:rdceg_colored}. The other situations ($s_{16}$, $s_{21}$, $s_{26}$ and $s_{33}$) appear to cause a large reduction in the log marginal likelihood score when they are left out of their stage. When more data becomes available, it should be used to verify the membership of these situations to their stage. Overall, the stages given by the model selection algorithm appear to be acceptable. 

Figure \ref{fig:loo_clusters} shows that the clusters chosen by the model selection algorithm are also reasonable. This can be seen from the fact that despite the use of the greedy search algorithm, the log marginal likelihood score is the highest when none of the edges in the cluster are left out and so in this sense is a local maximum. Also, the observed mean of the holding times on the edge that has been left out is always within one standard deviation of the posterior expectation of the cluster without that edge. However, we note here that the variance of the holding time distributions is quite high but overall, we find the stages and clusters created by the model selection algorithm to be within an acceptable range.  